\documentclass[submission, Phys]{SciPost}

\allowdisplaybreaks[1]

\hypersetup{
    colorlinks,
    linkcolor={red!50!black},
    citecolor={blue!50!black},
    urlcolor={blue!80!black}
}

\usepackage[bitstream-charter]{mathdesign}
\DeclareSymbolFont{usualmathcal}{OMS}{cmsy}{m}{n}
\DeclareSymbolFontAlphabet{\mathcal}{usualmathcal}

\usepackage{bm}
\usepackage{amsmath}
\usepackage{color}
\usepackage{bbm}
\usepackage{array}
\usepackage{graphicx}

\newcommand* {\vek}[1]{{\bm{\mathrm{#1}}}}
\newcommand* {\kk}{\vek{k}}
\newcommand* {\kkop}{\hat{\vek{k}}}

\newcommand* {\rr}{\vek{r}}
\newcommand* {\ee}{\mathrm{e}}
\newcommand{\sgn}{\mathrm{sgn}}

\let\myRe\Re
\let\myIm\Im
\renewcommand{\Re}{\myRe\mathrm{e}\,}
\renewcommand{\Im}{\myIm\mathrm{m}\,}

\begin{document}

\begin{center}
{\Large \textbf{Andreev bound states at boundaries of polarized
2D Fermi superfluids with \textit{s}-wave pairing and spin-orbit
coupling}}
\end{center}

\begin{center}
Kadin Thompson\textsuperscript{1},
Joachim Brand\textsuperscript{2} and
Ulrich Z\"ulicke\textsuperscript{1 $\star$}
\end{center}

\begin{center}
{\bf 1} Dodd-Walls Centre for Photonic and Quantum Technologies,
\\ School of Chemical and Physical Sciences, Victoria University
of Wellington, \\ PO Box 600, Wellington 6140, New Zealand
\\
{\bf 2} Dodd-Walls Centre for Photonic and Quantum Technologies,
Centre for Theoretical Chemistry and Physics, New Zealand
Institute for Advanced Study, Massey University,\\ Private Bag
102904, North Shore, Auckland 0745, New Zealand
\\[0.2cm]
${}^\star$ {\small \sf uli.zuelicke@vuw.ac.nz}
\end{center}

\begin{center}
\today
\end{center}


\section*{Abstract}
{\bf
A topological superfluid phase characterized by an emergent
chiral-\textit{p}-wave pair potential is expected to form in a
two-dimensional Fermi superfluid subject to \textit{s}-wave
pairing, spin-orbit coupling and a large-enough Zeeman
splitting. Andreev bound states appear at phase boundaries,
including Majorana zero modes whose existence is assured by the
bulk-boundary correspondence principle. Here we study the
physical properties of these subgap-energy bound states at
step-like interfaces using the spin-resolved
Bogoliubov--de$\,$Gennes mean-field formalism and assuming small
spin-orbit coupling. Extending a recently developed
spin-projection technique based on Feshbach partitioning
\href{http://dx.doi.org/10.21468/SciPostPhys.5.2.016}{[SciPost
Phys.\ \textbf{5}, 016 (2018)]} combined with the Andreev
approximation allows us to obtain remarkably simple analytical
expressions for the bound-state energies as well as the
majority and minority spin components of their wave functions.
Besides the vacuum boundary, where a majority-spin Majorana
excitation is encountered, we also consider the boundary
between the topological and a nontopological superfluid phase
that can appear in a coexistence scenario due to the
first-order topological phase transition predicted for this
system. At this superfluid-superfluid interface, we find a
localized chiral Majorana mode hosted by the minority-spin
sector. Our theory further predicts majority-spin subgap-energy
bound states similar to those found at a Josephson junction
between same-chirality \textit{p}-wave superfluids. Their
presence affects the Majorana mode due to a coupling of
minority and majority spin sectors only in the small energy
range where their spectra overlap. Our results may inform
experimental efforts aimed at realizing and characterizing
unconventional Majorana quasiparticles.
}

\vspace{10pt}
\noindent\rule{\textwidth}{1pt}
\tableofcontents\thispagestyle{fancy}
\noindent\rule{\textwidth}{1pt}
\vspace{10pt}

\section{Introduction}
\label{sec:intro}

The prospect of using condensed-matter
realizations~\cite{Alicea2012,Beenakker2013} of Majorana
fermions~\cite{Wilczek2009,Elliott2015} for fault-tolerant
quantum computation~\cite{DasSarma2015,Lia2018} has spurred
intense efforts aimed at creating topological
superfluids~\cite{Beenakker2016,Sato2017,Flensberg2021}. One
of the promising proposals~\cite{Zhang2008} is based on driving
two-dimensional (2D) Fermi systems with attractive
\textit{s}-wave interaction \cite{Levinsen2015} and
spin-orbit coupling~\cite{Galitski2013,Meng2016} into a
topological-superfluid (TSF) phase by increasing the Zeeman
spin-splitting energy $h$ above a critical value, which is given
by~\cite{Sau2010,Alicea2010,Sato2010}
\begin{equation}\label{eq:hcrit}
h_\mathrm{c} = \sqrt{\mu^2 + |\Delta|^2}
\end{equation}
in terms of the superfluid's chemical potential $\mu$ and
\textit{s}-wave pair-potential magnitude $|\Delta|$. In the TSF
regime, the majority-spin (here: spin-$\uparrow$) degrees of
freedom govern the system's low-energy properties, exhibiting
characteristics of a spinless chiral 2D \textit{p}-wave
superfluid~\cite{Kallin2016} that is known to have topologically
protected zero-energy excitations in vortices~\cite{Kopnin1991,
Volovik1999,Read2000,Ivanov2001,Tewari2007} and at its
boundary~\cite{Honerkamp1998,Matsumoto1999,Furusaki2001,
Stone2004,Mizushima2008,Fu2008,Sauls2011}. However, the
necessarily incomplete quenching of the minority-spin (i.e., 
spin-$\downarrow$) sector can affect the system's physical
properties in such a way as to spoil the perfect congruence with
an ideal chiral-\textit{p}-wave superfluid~\cite{Brand2018,
Thompson2020}. Understanding the effect this has on the
microscopic properties of low-energy excitations emerging at the
edge of a 2D TSF is one of the main purposes of the present
work. The results we obtain here illuminate and extend insights
gained from previous numerical studies~\cite{Sato2009,Holst2022}
and related earlier work~\cite{Ghosh2010}. Furthermore, the
possibility to have situations where the system splits into
coexisting TSF and nontopological-superfluid (NSF)
parts~\cite{Zhou2011,Liu2012,Thompson2021} motivates our study
of subgap excitations localized at a TSF-NSF interface.

\begin{figure}[t]
\centerline{%
\includegraphics[width=0.9\textwidth]{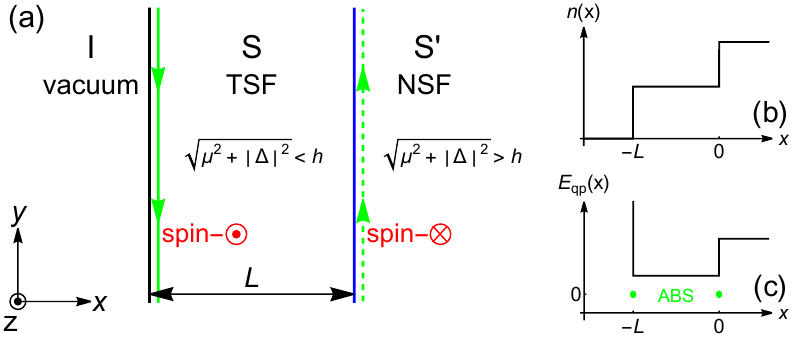}
}%
\caption{\label{fig:ISSprime}%
(a)~Schematic illustration of the ISS$^\prime$ model system
focused on in this work. The solid and dashed green lines
indicate the chiral Majorana edge modes arising at the boundary
of the topological superfluid (TSF) with vacuum (I) and the
nontopological superfluid (NSF), respectively, for which both
spin and propogation direction turn out to be opposite.
(b)~Density profile for a typical physical realization of the
considered ISS$^\prime$ system. A 2D Fermi gas undergoing a
first-order topological transition~\cite{Thompson2021} splits
up into coexisting TSF and NSF parts having the same chemical
potential but different densities and pair-potential magnitudes.
The TSF region has the smaller density and smaller
pair-potential magnitude~\cite{Thompson2021}. (c)~The spatial
variation of the minimum energy $E_\mathrm{qp}$ required for
quasiparticle excitations is shown as the black solid curve. We
assume $E_\mathrm{qp}=\infty$ in the I region so that no
quasiparticles are present there. Green dots indicate the
minimum energy $E_\mathrm{qp}=0$ of the interface-localized
Andreev bound states (ABS) that constitute the Majorana edge
modes.}
\end{figure}

The basic setup of our system of interest is shown schematically
in Fig.~\ref{fig:ISSprime}(a). It constitutes an ISS$^\prime$
hybrid system, where I stands for an insulator --- a region of
space with practically unreachable high (quasi)particle
excitation energies---, while S and S$^\prime$ are adjacent
regions of space in which a 2D Fermi superfluid subject to
\textit{s}-wave pairing, spin-orbit coupling and Zeeman spin
splitting is present in the TSF and NSF phase, respectively. To
be specific, we assume that the two-particle attraction,
spin-orbit coupling and Zeeman splitting are uniform across all
these regions, and that the chemical potential is also the same
throughout. In contrast, the \textit{s}-wave superfluid pair
potential $\Delta(\rr) \equiv\Delta(x)$ is considered to be a
function of the coordinate $x$ in the direction perpendicular to
the SS$^\prime$ interface. Thus, while the IS boundary arises
from an ordinary single-particle confining potential, the
SS$^\prime$ interface is due to different superfluid order
parameters being present in the S and S$^\prime$ regions. In
this, our envisioned realization of the TFS-NSF hybrid system
differs from the one focused on in a previous
study~\cite{Setiawan2017} where the pair-potential magnitude was
assumed to be the same in the S and S$^\prime$ regions but their
chemical potentials were different. The intrinsically broken
time-reversal symmetry due to the finite Zeeman energy
distinguishes our system of interest also from the one
considered in Ref.~\cite{Rodriguez2022} where the TSF is a
time-reversal invariant topological superfluid (TRITOPS).

The type of ISS$^\prime$ system focused on in the present work
can be realized experimentally, e.g., by a Fermi system that has
split into coexisting TSF and NSF phases. Such a phase
coexistence was suggested to occur in trapped 2D Fermi
gases~\cite{Zhou2011,Liu2012} where the external potential
causes spatially varying $\mu$ and $\Delta$. As a result,
$h_\mathrm{c}$ also becomes position-dependent, making it
possible for TSF and NSF regions to exist simultaneously within
a trap at fixed Zeeman energy $h$. Alternatively, TSF-NSF phase
coexistence arises in a 2D Fermi gas undergoing a first-order
topological transition~\cite{Thompson2021}. In this latter
scenario, which most closely resembles the situation
investigated in the following, thermodynamics dictates the
homogeneous-density values in the adjacent TSF and NSF regions
and, thus, their relative size. The fact that the TSF phase
generically has lower density~\cite{Thompson2021} implies that
it will be positioned near the system boundary and, as a result,
the typical density profile of the ISS$^\prime$ system has the
form sketched in Fig.~\ref{fig:ISSprime}(b). To ensure that the
superfluid phases in the S and S$^\prime$ regions are on the BCS
side of the BCS-BEC crossover~\cite{Randeria1990}, $\mu > 0$ is
assumed throughout this work.

Obtaining the fully self-consistent spatial profile of the pair
potential in a general ISS$^\prime$ system will usually require
numerical calculations as described, e.g., in
Ref.~\cite{Spuntarelli2010}. Regardless of any details, however,
it can be expected that the homogeneous-system forms of pair
potentials in the individual S and S$^\prime$ phases are
recovered sufficiently far away from their boundaries and the
SS$^\prime$ interface. We therefore assume the width $L$ of the
S region to be large [see illustration in
Fig.~\ref{fig:ISSprime}(a)] and make the \textit{Ansatz}
\begin{equation}\label{eq:DeltaX}
\Delta(x) = \big| \Delta^{(\mathrm{S})} \big|\, \ee^{i
\varphi^{(\mathrm{S})}} \, \Theta( x + L)\Theta(-x) + \big|
\Delta^{(\mathrm{S}^\prime)} \big| \, \ee^{i
\varphi^{(\mathrm{S'})}} \, \Theta(x) \quad .
\end{equation}
Here $\Theta(\cdot)$ denotes the Heaviside step function,
$\varphi^{(R)} \equiv \mathrm{arg}\big(\Delta^{(R)}\big)$ is the
phase of the complex \textit{s}-wave pair potential in region
$R\in \{\mathrm{S}, \mathrm{S'}\}$, and we assumed without loss
of generality that the SS$^\prime$ interface is located at
$x=0$. The fact that the S region is a TSF and the S$^\prime$
region is a NSF, respectively, implies the relation
\begin{equation}\label{eq:TSFtoNSF}
\sqrt{\mu^2 + \big| \Delta^{(\mathrm{S})} \big|^2} < h <
\sqrt{\mu^2 + \big| \Delta^{(\mathrm{S}^\prime)} \big|^2} \,\, ,
\end{equation}
which requires $\mu < h$ and $\big|\Delta^{(\mathrm{S})}\big| <
\big|\Delta^{(\mathrm{S}^\prime)}\big|$. To provide a measure
for how deep into the TSF (the NSF) phase the superfluid
occupying the S (the S$^\prime$) region is, we introduce
\begin{equation}\label{eq:junctDeltaCri}
\Delta_\mathrm{c} = \sqrt{h^2 - \mu^2}
\end{equation}
as the critical \textit{s}-wave pair-potential magnitude for the
hybrid system and consider situations where $\Delta_\mathrm{c} -
\big|\Delta^{(\mathrm{S})}\big| \gtrsim \Delta_\mathrm{c}$ (where
$\big|\Delta^{(\mathrm{S'})}\big| - \Delta_\mathrm{c} \sim
\Delta_\mathrm{c}$) to be well-developed TSFs (NSFs).

Due to the influence of spin-orbit coupling and Zeeman spin
splitting, the minimum quasi\-particle-excitation energy
$E_\mathrm{qp}$ is not equal to the modulus $|\Delta|$ of the
\textit{s}-wave pair potential in each respective region. A more
detailed discussion of how $E_\mathrm{qp}$ depends on system
parameters is provided in Sec.~\ref{sec:formal} below [see
Eq.~(\ref{eq:MinQP}) and following paragraph]. For our purposes,
we limit consideration to the case illustrated by
Fig.~\ref{fig:ISSprime}(c) where $E_\mathrm{qp}^{(\mathrm{S})}<
E_\mathrm{qp}^{(\mathrm{S}^\prime)}$, as this corresponds to the
situation encountered typically in TSF-NSF hybrid systems
arising from phase coexistence during a first-order topological
transition~\cite{Thompson2021}. Our study focuses on Andreev
bound states that are localized at the individual interfaces and
whose energies therefore satisfy $|E| < 
E_\mathrm{qp}^{(\mathrm{S})}$. The S region also hosts extended
Andreev bound states with $E_\mathrm{qp}^{(\mathrm{S})} < |E| <
E_\mathrm{qp}^{(\mathrm{S}^\prime)}$ that are confined by both
interfaces and whose energies depend on $L$, but these are not
further investigated here. The evanescent nature of
interface-localized Andreev bound states makes it possible for
the IS and SS$^\prime$ interfaces to be considered independently
of each other, as long as their distance $L$ is much larger than
the spatial decay length for the evanescent-quasiparticle
amplitudes~\cite{Sauls2018}. We assume this to be the case in
the following.

This article is organized as follows. The basic mathematical
formalism for obtaining evanescent Andreev bound states in
inhomogeneous 2D Fermi superfluids with spin-orbit coupling and
Zeeman spin splitting is developed in Sec.~\ref{sec:formal}.
We first introduce the spin-resolved version of
Bogoliubov-de~Gennes theory on which our approach is based.
Following that, the Feshbach-partitioning technique used
previously~\cite{Brand2018} for accurate calculation of
quasiparticle dispersions and associated thermodynamic
properties of TSF and NSF phases is adapted and extended to
obtain Andreev bound states. In particular, we show in
Sec~\ref{sec:projTech} how the excitations of the homogeneous
superfluid can be faithfully represented in terms of
quasiparticles of chiral-\textit{p}-wave superfluids associated
with a fixed spin projection. The form of the specific
\textit{Ans\"atze}\/ and matching conditions used for treating
the IS and SS$^\prime$ interfaces are then given in
Secs.~\ref{sec:ISform} and \ref{sec:SSPform}, respectively. To
enable a physically realistic description of these hybrid
systems, our theory accounts for the residual coupling between
the effective chiral-\textit{p}-wave superfluids realized within
opposite-spin subsectors. Results obtained when applying this
formalism to the edge of an individual TSF or NSF are presented
in Sec~\ref{sec:resSurf}, and the Andreev bound states localized
at the TSF-NSF interface are discussed in
Sec.~\ref{sec:resInter}. In both these cases, we obtain
analytical results and use these to juxtapose the physical
properties of Andreev bound states in our system of interest
with those found previously for simpler systems. Our
conclusions, together with a discussion and outlook on
experimental ramifications of our results, are given in the
final section \ref{sec:concl}. As the subgap excitations at
Josephson junctions between spinless chiral-\textit{p}-wave
superfluids provide an instructive reference point to discuss
our predictions, we present relevant background information in
Appendix~\ref{app:chiPwave}. In the process, we also generalize
previous results~\cite{Ho1984,Barash2001,Kwon2004,Samokhin2012}
to the situation where the chiral-\textit{p}-wave pair-potential
magnitudes are different on opposite sides of the
junction.\footnote{The effect of unequal pair-potential
magnitudes on Andreev-bound-state spectra is implicit in
theoretical treatments of \textit{d}-wave Josephson
junctions~\cite{Tanaka1996,Kashiwaya2000} and has also been
considered for junctions of \textit{s}-wave
superfluids}~\cite{Presilla2017}. A brief pedagogical
introduction discussing the emergence and properties of Majorana
edge modes at the boundary of chiral-\textit{p}-wave superfluids
is provided in Appendix~\ref{app:Majorana}. Readers who already
possess some relevant background knowledge and prefer learning
first about our predictions rather than the formalism could skip
parts or all of Sec.~\ref{sec:formal} on their first reading.

\section{Formal description of interface-localized Andreev bound
states}\label{sec:formal}

Quasiparticle states and excitation energies of a 2D Fermi
superfluid with \textit{s}-wave pairing, spin-orbit coupling and
Zeeman spin splitting are obtained as solutions of the
spin-resolved Bogoliubov--de~Gennes (BdG)
equation~\cite{deGennes1989,Ketterson1999}
\begin{equation}\label{eq:4x4ham}
\begin{pmatrix}
\epsilon_{\hat{\kk}} - h - \mu & 0 & \vek{\lambda}\cdot
\hat{\kk} & -\Delta(\rr) \\[0.1cm]
0 & -\epsilon_{\hat{\kk}} + h + \mu & \Delta^\ast(\rr) &
\vek{\lambda}^\ast\cdot\hat{\kk} \\[0.1cm]
\vek{\lambda}^\ast\cdot\hat{\kk} & \Delta(\rr) &
\epsilon_{\hat{\kk}} + h - \mu & 0 \\[0.1cm]
-\Delta^*(\rr) & \vek{\lambda}\cdot\hat{\kk} & 0 &
-\epsilon_{\hat{\kk}} - h + \mu
\end{pmatrix} \begin{pmatrix}
u_\uparrow(\rr) \\[0.1cm] v_\uparrow(\rr) \\[0.1cm]
u_\downarrow(\rr) \\[0.1cm] v_\downarrow(\rr) \end{pmatrix} = E
\, \begin{pmatrix} u_\uparrow(\rr) \\[0.1cm] v_\uparrow(\rr)
\\[0.1cm] u_\downarrow(\rr) \\[0.1cm] v_\downarrow(\rr)
\end{pmatrix} \quad ,
\end{equation}
which contains the \textit{s}-wave pair potential $\Delta(\rr)$,
chemical potential $\mu$ and Zeeman energy $h$ that have already
been introduced above. Furthermore, `$\ast$' denotes complex
conjugation, $\hat{k} \equiv (-i \partial_x, -i \partial_y)$
is the 2D wave vector in position-space representation,
$\epsilon_{\hat{\kk}} \equiv \hbar^2 (\hat{k}_x^2 + \hat{k}_y^2)
/(2 m)$ corresponds to the single-particle-energy dispersion
(assumed to be parabolic for simplicity), and spin-orbit
coupling is embodied in the vector $\vek{\lambda} = \lambda\,
(i\, , 1)$ whose particular form represents Rashba spin-orbit
coupling~\cite{Bychkov1984,Bihlmayer2015,Manchon2015}. Without
loss of generality, we assume $h>0$ and $\lambda>0$ throughout
this work. Considering a piecewise-constant pair potential
$\Delta(\rr) \equiv \Delta(x)$ as per Eq.~(\ref{eq:DeltaX}), a
general solution of the BdG equation (\ref{eq:4x4ham}) for the
entire ISS$^\prime$ hybrid system is of the form
\begin{equation}\label{eq:GenAnsatze}
\begin{pmatrix} u_\uparrow(\rr) \\[0.1cm] v_\uparrow(\rr)
\\[0.1cm] u_\downarrow(\rr) \\[0.1cm] v_\downarrow(\rr)
\end{pmatrix} = \left[ \begin{pmatrix}
u_{\uparrow}^{(\mathrm{I})}(x) \\[0.1cm]
v_{\uparrow}^{(\mathrm{I})}(x) \\[0.1cm]
u_{\downarrow}^{(\mathrm{I})}(x) \\[0.1cm]
v_{\downarrow}^{(\mathrm{I})}(x) \end{pmatrix} \Theta(- x - L)
+ \begin{pmatrix} u_{\uparrow}^{(\mathrm{S})}(x) \\[0.1cm]
v_{\uparrow}^{(\mathrm{S})}(x) \\[0.1cm]
u_{\downarrow}^{(\mathrm{S})}(x) \\[0.1cm]
v_{\downarrow}^{(\mathrm{S})}(x) \end{pmatrix} \Theta(x + L)
\Theta(-x) + \begin{pmatrix}
u_{\uparrow}^{(\mathrm{S}^\prime)}(x) \\[0.1cm]
v_{\uparrow}^{(\mathrm{S}^\prime)}(x) \\[0.1cm]
u_{\downarrow}^{(\mathrm{S}^\prime)}(x) \\[0.1cm]
v_{\downarrow}^{(\mathrm{S}^\prime)}(x)\end{pmatrix} \Theta(x)
\right] \ee^{i k_y y} \, ,
\end{equation}
which is a combination of Nambu spinors that are solutions of
(\ref{eq:4x4ham}) within the individual I, S, and S$^\prime$
regions for fixed energy $E$ and wave-vector component $k_y$,
joined smoothly across the two interfaces. Based on our
assumption that the 2D Fermi superfluid is homogeneous within a
given region $R\in\{\mathrm{I},\mathrm{S},\mathrm{S'}\}$, we can
write the $x$-dependent part of the Nambu four-spinor pertaining
to $R$ as a superposition of plane-wave eigenstates of the $4
\times 4$ BdG equation (\ref{eq:4x4ham}) with constant $\Delta
(\rr)\equiv \Delta^{(R)}$ [see Eq.~(\ref{eq:DeltaX}):
$\Delta^{(\mathrm{I})}=0$, $\Delta^{(\mathrm{S})}=\big|
\Delta^{(\mathrm{S})}\big|\exp\big( i \varphi^{(\mathrm{S})}
\big)$ and $\Delta^{(\mathrm{S}^\prime)} = \big|
\Delta^{(\mathrm{S'})}\big|\, \exp\big( i\varphi^{(\mathrm{S'})}
\big)$];
\begin{equation}\label{eq:genPWans}
\begin{pmatrix} u_{\uparrow}^{(R)}(x) \\[0.1cm]
v_{\uparrow}^{(R)}(x) \\[0.1cm] u_{\downarrow}^{(R)}(x)
\\[0.1cm] v_{\downarrow}^{(R)}(x) \end{pmatrix} = \sum_\zeta\,
a^{(R)}_\zeta\,\,\, \begin{pmatrix} u_{\uparrow\kk_\zeta^{(R)}}
\\[0.1cm] v_{\uparrow\kk_\zeta^{(R)}} \\[0.1cm] u_{\downarrow
\kk_\zeta^{(R)}}\\[0.1cm] v_{\downarrow\kk_\zeta^{(R)}}
\end{pmatrix}\,\,\, \ee^{i\, k_\zeta^{(R)}\, x} \quad .
\end{equation}
Here $\zeta$ labels the different wave vectors $\kk^{(R)}_\zeta
\equiv \big( k_\zeta^{(R)}, k_y \big)$ for which eigenstates of
(\ref{eq:4x4ham}) having energy $E$ exist, $\big( u_{\uparrow
\kk}, v_{\uparrow\kk}, u_{\downarrow\kk}, v_{\downarrow\kk}
\big)^T$ is the Nambu four-spinor associated with the plane-wave
eigenstate of (\ref{eq:4x4ham}) with wave vector $\kk$ and
energy $E$ [see Eq.~(\ref{eq:4x4plane}) below for a more
explicit form of the full eigenstate], and the $a^{(R)}_\zeta$
are complex coefficients that are fixed by the requirement to
satisfy the boundary conditions for region $R$.

In principle, the energy dispersions and associated eigenstates
for a homogeneous 2D Fermi superfluid described by the $4\times
4$ BdG equation (\ref{eq:4x4ham}) with constant $\Delta(\rr)
\equiv \Delta$ are available in closed-analytical
form~\cite{Zhou2011,Thompson2020}. However, the matching
conditions for superpositions of these exact Nambu four-spinor
solutions from each region at the two interfaces generate a
complicated system of equations that is not straightforwardly
tractable, even numerically. To circumvent this complexity and
gain useful analytical insight, we use approximate expressions
for energies and Nambu spinors arising from an accurate
projection technique developed in Ref.~\cite{Brand2018}. We
describe the fundamentals of this approach in the following
subsection~\ref{sec:projTech}. Following that, the application
of the formalism to obtain Andreev bound states at the
individual IS and SS$^\prime$ interfaces is discussed in
subsections \ref{sec:ISform} and \ref{sec:SSPform},
respectively.

\subsection{Approximate spin-projected description of
homogeneous-system excitations}\label{sec:projTech}

A homogeneous polarized 2D Fermi superfluid with spin-orbit
coupling is described by the $4\times 4$ BdG equation
(\ref{eq:4x4ham}) with constant $\Delta(\rr)\equiv \Delta$.
Using the plane-wave \textit{Ansatz}
\begin{equation}\label{eq:4x4plane}
\begin{pmatrix} u_\uparrow(\rr) \\[0.1cm] v_\uparrow(\rr)
\\[0.1cm] u_\downarrow(\rr) \\[0.1cm] v_\downarrow(\rr)
\end{pmatrix} = \begin{pmatrix} u_{\uparrow\kk} \\[0.1cm]
v_{\uparrow\kk} \\[0.1cm] u_{\downarrow\kk} \\[0.1cm]
v_{\downarrow\kk} \end{pmatrix} \,\,\, \ee^{i\, \kk\cdot\rr}
\end{equation}
with 2D wave vector $\vek{k} \equiv (k_x, k_y)$, the BdG
equation (\ref{eq:4x4ham}) is straightforwardly solved. In
particular, the homogeneous-system quasi\-particle-energy
spectrum is found to have four dispersion
branches \cite{Yi2011,Zhou2011};
\begin{subequations}\label{eq:fullSpec}
\begin{align}\label{eq:fullSpecSm}
E_{\kk<\eta} &= \eta\, \sqrt{\left( \epsilon_\kk - \mu \right)^2
+ |\Delta|^2 + h^2 + \lambda^2\, \kk^2 - 2 \sqrt{\left(
\epsilon_\kk - \mu \right)^2 \left( h^2 + \lambda^2\, \kk^2
\right) + |\Delta|^2 h^2}} \quad , \\
E_{\kk>\eta} &= \eta\, \sqrt{\left( \epsilon_\kk -\mu \right)^2
+ |\Delta|^2 + h^2 + \lambda^2\, \kk^2 + 2 \sqrt{\left(
\epsilon_\kk - \mu \right)^2 \left( h^2 + \lambda^2\, \kk^2
\right) + |\Delta|^2 h^2}} \quad ,
\end{align}
\end{subequations}
with $\eta \in \{+, -\}$ distinguishing positive-energy and
negative-energy states.

\begin{figure}[t]
\centerline{%
\includegraphics[width=0.8\textwidth]{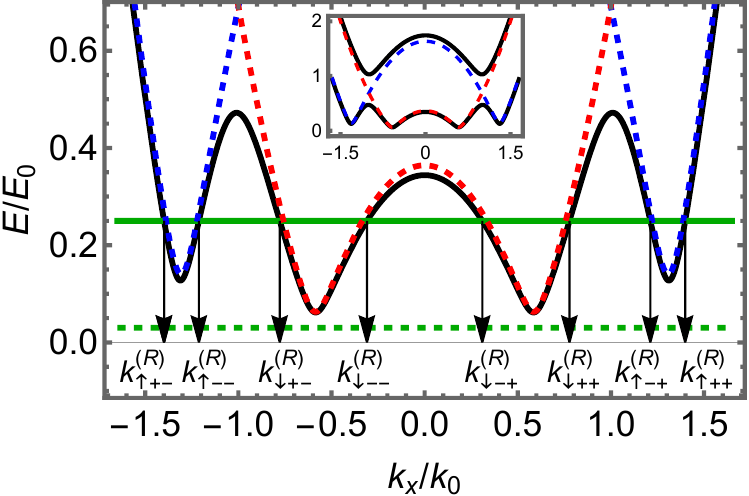}
}%
\caption{\label{fig:WaveVecs}%
Approximate description of the low-energy quasiparticle
dispersion in the nontopological-superfluid (NSF) phase. The
black solid curve shows the lower positive-energy branch
$E_{\kk < +}$ [see explicit expression given in
Eq.~(\ref{eq:fullSpecSm})] of quasiparticle energies obtained
from diagonalizing the Hamiltonian in Eq.~(\ref{eq:4x4ham}) for
$k_y = 0$ and with the parameters $\mu/E_0 = 1.00$, $|\Delta|/
E_0 = 0.30$ , $h/E_0 = 0.70$, $\lambda k_0/E_0 = 0.25$. Here
$k_0$ and $E_0$ are arbitrary wave-number and energy units
related via $E_0 =\hbar^2 k_0^2/(2m)$. For reference, the inset
plots both positive-energy branches of the
quasiparticle-excitation spectrum [$E_{\kk < +}$ and $E_{\kk >
+}$ as per Eqs.~(\ref{eq:fullSpec})] as black solid curves. The
blue dashed (red dashed) curve is the approximate energy
dispersion $E_{\vek{k}\uparrow +}$ ($E_{\vek{k}\downarrow +}$)
from Eq.~(\ref{eq:lowEdisp}). Intersection of the dispersions
with a fixed-energy value $E$ above the quasiparticle-excitation
gap (solid green horizontal line) defines wave-vector components
$k^{(R)}_{\sigma\tau\alpha}$ associated with quasiparticle
excitations moving perpendicular to the interfaces in region
$R$, for which approximate analytical expressions are given in
Eq.~(\ref{eq:ProjWaveVecs}). For a value of $E$ below the
quasiparticle-excitation gap (dashed green horizontal line),
these wave numbers are complex-valued.}
\end{figure}

\subsubsection{Formally exact 2$\times$2 projection}

The inset of Fig.~\ref{fig:WaveVecs} shows a plot of the
positive-energy branches $E_{\kk<+}$ and $E_{\kk>+}$. It is
possible to represent the low-energy part of these dispersions
quite faithfully in terms of a set of spin-projected dispersion
relations (shown as blue and red dashed curves in
Fig.~\ref{fig:WaveVecs}). This approach is motivated by the
observation that, for small-enough pair-potential magnitude and
not-too-large spin-orbit-coupling strength, as compared with the
spin-polarizing Zeeman energy,\footnote{Specifically, $|\Delta
|^2\ll \mu h$ and $\lambda^2 \ll \hbar^2 h/m$ are the required
conditions. The first of these typically applies in the BCS
regime for \textit{s}-wave pairing, but the approach turns out
to also describe the TSF in the BEC regime~\cite{Thompson2020}.}
the plane-wave Nambu eigenspinors (\ref{eq:4x4plane}) that are
the solutions of (\ref{eq:4x4ham}) for the homogeneous system
have generally either dominating spin-$\uparrow$ character or
dominating spin-$\downarrow$ character, i.e., they satisfy
\begin{equation}
| u_{\sigma\kk} |^2 + | v_{\sigma\kk}|^2 \gg | u_{\bar{\sigma}
\kk} |^2 + | v_{\bar{\sigma}\kk} |^2 \quad ,
\end{equation}
with $\bar{\sigma}$ being the opposite of $\sigma$ [i.e.,
$\bar{\sigma} =\,\, \downarrow$ ($\uparrow$) if  $\sigma =\,\,
\uparrow$ ($\downarrow$)]. The formal rewriting of the
spin-resolved BdG equation (\ref{eq:4x4ham}) for constant
$\Delta(\rr)\equiv\Delta$ in plane-wave representation in terms
of the two $2\times 2$ equations 
\begin{subequations}
\begin{align} \label{eq:2x2BdG}
\left( \mathcal{H}_{\sigma\sigma} - \mathcal{H}_{\sigma
\bar{\sigma}} \left[ \mathcal{H}_{\bar{\sigma}\bar{\sigma}} - E
\, \mathbbm{1} \right]^{-1} \, \mathcal{H}_{\bar{\sigma}\sigma}
\right) \, \begin{pmatrix} u_{\sigma\kk} \\ v_{\sigma\kk}
\end{pmatrix} =&\, E \, \begin{pmatrix} u_{\sigma\kk} \\
v_{\sigma\kk} \end{pmatrix} \quad , \\[0.2cm] \label{eq:2x2def}
\begin{pmatrix} u_{\bar{\sigma}\kk} \\ v_{\bar{\sigma}\kk}
\end{pmatrix} =&\, - \left[
\mathcal{H}_{\bar{\sigma}\bar{\sigma}} - E \, \mathbbm{1}
\right]^{-1} \, \mathcal{H}_{\bar{\sigma}\sigma} \,
\begin{pmatrix} u_{\sigma\kk} \\ v_{\sigma\kk} \end{pmatrix}
\quad ,
\end{align}
\end{subequations}
where
\begin{subequations}
\begin{eqnarray}
& \mathcal{H}_{\uparrow\uparrow} = \begin{pmatrix}
\epsilon_\kk - h - \mu & 0 \\ 0 & -\epsilon_\kk + h + \mu
\end{pmatrix} \quad , \quad
\mathcal{H}_{\downarrow\downarrow} = \begin{pmatrix}
\epsilon_\kk + h - \mu & 0 \\ 0 & -\epsilon_\kk - h + \mu
\end{pmatrix} & \,\, , \\[0.2cm]
& \mathcal{H}_{\uparrow\downarrow} \equiv \left(
\mathcal{H}_{\downarrow\uparrow} \right)^\dagger =
\begin{pmatrix} \vek{\lambda}\cdot\kk & -\Delta \\[0.1cm]
\Delta^* & \vek{\lambda}^*\cdot\kk \end{pmatrix} & \,\, ,
\end{eqnarray}
\end{subequations}
forms the basis for an approximate treatment where Nambu
four-spinor solutions of (\ref{eq:4x4ham}) for the homogeneous
system are represented in terms of their respective large
amplitudes $u_{\sigma\kk}$ and $v_{\sigma\kk}(\rr)$.
Specializing the more general expressions from
Ref.~\cite{Brand2018} to the case of weak spin-orbit coupling
$\lambda |\kk| \ll h$, we find
\begin{subequations}
\begin{align}\label{eq:2x2BdGup}
\begin{pmatrix}
\xi_{\kk \uparrow} & - \frac{\vek{\lambda} \cdot \kk}{h} \,
\Delta \\[0.1cm] - \frac{\vek{\lambda^\ast}\cdot \kk}{h} \,
\Delta^\ast & - \xi_{\kk\uparrow}\end{pmatrix} \begin{pmatrix}
u_{\uparrow\kk} \\[0.1cm] v_{\uparrow\kk} \end{pmatrix} &=
E_{\kk\uparrow}\, \begin{pmatrix} u_{\uparrow\kk} \\[0.1cm]
v_{\uparrow\kk} \end{pmatrix} \quad , \\[0.2cm]
\label{eq:2x2downUp}
\begin{pmatrix} u_{\downarrow\kk} \\[0.1cm] v_{\downarrow\kk}
\end{pmatrix} &= \begin{pmatrix} -\frac{\vek{\lambda^*}\cdot
\kk}{2h} & -\frac{\Delta}{2h} \\[0.1cm] -\frac{\Delta^*}{2h} &
\frac{\vek{\lambda}\cdot\kk}{2h} \end{pmatrix} \begin{pmatrix}
u_{\uparrow\kk} \\[0.1cm] v_{\uparrow\kk} \end{pmatrix}
\end{align}
\end{subequations}
for the large-spin-$\uparrow$ spinors, and similarly
\begin{subequations}
\begin{align}\label{eq:2x2BdGdown}
\begin{pmatrix}\xi_{\kk \downarrow} & - \frac{\vek{\lambda^*}
\cdot \kk}{h} \, \Delta \\[0.1cm] - \frac{\vek{\lambda} \cdot
\kk}{h} \, \Delta^\ast & - \xi_{\kk\downarrow} \end{pmatrix}
\begin{pmatrix} u_{\downarrow\kk} \\[0.1cm] v_{\downarrow\kk}
\end{pmatrix} &= E_{\kk\downarrow}\, \begin{pmatrix}
u_{\downarrow\kk} \\[0.1cm] v_{\downarrow\kk} \end{pmatrix}
\quad , \\[0.2cm] \label{eq:2x2upDown} \begin{pmatrix}
u_{\uparrow\kk} \\[0.1cm] v_{\uparrow\kk} \end{pmatrix} &=
\begin{pmatrix} \frac{\vek{\lambda}\cdot\kk}{2h} &
-\frac{\Delta}{2h} \\[0.1cm] -\frac{\Delta^*}{2h} & 
-\frac{\vek{\lambda^*}\cdot\kk}{2h} \end{pmatrix}
\begin{pmatrix} u_{\downarrow\kk} \\[0.1cm] v_{\downarrow\kk}
\end{pmatrix}
\end{align}
\end{subequations}
for the large-spin-$\downarrow$ spinors. Here we introduced the
effective unpaired-quasiparticle energy 
\begin{equation}\label{eq:xizDef}
\xi_{\kk\sigma} = \frac{\hbar^2}{2 m_\sigma}\, \kk^2 +
\nu_\sigma - \mu \quad .
\end{equation}
Within our leading-order-in-$\lambda |\kk|/h$ approximation,
which amounts to keeping terms upto qua\-drat\-ic order in
$\lambda |\kk|/h$ in the effective $2\times 2$ BdG Hamiltonians
of Eqs.~(\ref{eq:2x2BdGup}) and (\ref{eq:2x2BdGdown}) but only
terms upto linear order in $\lambda |\kk|/h$ in the determining
relations (\ref{eq:2x2downUp}) and (\ref{eq:2x2upDown}) for the
small spinor amplitudes, we have
\begin{equation}\label{eq:mSigma}
m_\uparrow = \frac{m}{1 - \frac{m\, \lambda^2}{\hbar^2 h} \left(
1 + \frac{|\Delta|^2}{4h^2} \right)} \qquad , \qquad
m_\downarrow = \frac{m}{1 + \frac{m\, \lambda^2}{\hbar^2 h}
\left( 1 + \frac{|\Delta|^2}{4h^2} \right)}
\end{equation}
for the spin-dependent effective quasiparticle mass, and
\begin{equation}\label{eq:nuSigma}
\nu_\uparrow = - h + \frac{|\Delta|^2}{2h} \qquad , \qquad
\nu_\downarrow = h - \frac{|\Delta|^2}{2h}
\end{equation}
for the spin-dependent band-bottom shift entering
Eq.~(\ref{eq:xizDef}).

\subsubsection{Energy dispersions for weak spin-orbit coupling}

Diagonalization of (\ref{eq:2x2BdGup}) and (\ref{eq:2x2BdGdown})
yields the dominant-spin-$\sigma$-quasiparticle energy
dispersions
\begin{equation}\label{eq:lowEdisp}
E_{\vek{k}\sigma \eta} = \eta\,\sqrt{\xi_{\vek{k}\sigma}^2
+ \frac{\lambda^2 |\Delta|^2}{h^2}\, \vek{k}^2} \equiv \eta\,
\sqrt{\left( \xi_{\vek{k}\sigma} + \frac{\hbar^2}{m_\sigma}\,
k_{\Delta\sigma}^2\right)^2 + \bar{\Delta}_\sigma^2} \quad ,
\end{equation}
where we introduced the abbreviations
\begin{equation}\label{eq:pAbbrev}
\bar{\Delta}_\sigma = \Delta_\sigma \, \sqrt{1 - \frac{k_{\Delta
\sigma}^2}{k_{\mathrm{F}\sigma}^2}} \,\,\, , \,\,\,
\Delta_\sigma = \frac{\lambda\, k_{\mathrm{F}\sigma}}{h}\,
|\Delta| \,\,\, , \,\,\, k_{\mathrm{F}\sigma} = \sqrt{\frac{2
m_\sigma}{\hbar^2} \left( \mu - \nu_\sigma \right)} \,\,\, ,
\,\,\, k_{\Delta\sigma} = \frac{m_\sigma}{\hbar^2}\,
\frac{\lambda\, |\Delta|}{h} \,\,\, .
\end{equation}
Although $|k_{\Delta\sigma}|\ll |k_{\mathrm{F}\sigma}|$ holds
typically, we have retained $k_{\Delta\sigma}$-dependent terms
in the above expressions for the sake of formal consistency. The
dispersions (\ref{eq:lowEdisp}) formally coincide with the
quasiparticle energies obtained for \textit{p}-wave
pairing~\cite{Sigrist1991} and reproduce the low-energy part of
the true quasiparticle-excitation spectrum. See
Fig.~\ref{fig:WaveVecs} for a comparison. Straightforward
calculation yields the minimum excitation energy of the
spin-$\sigma$ branch as~\cite{Setiawan2015}
\begin{equation}\label{eq:MinQP}
\min \big| E_{\vek{k}\sigma \eta} \big| = \left\{
\begin{array}{cll} |\bar{\Delta}_\sigma|\,\, , & \mbox{occurring
where $|\vek{k}| = \sqrt{k_{\mathrm{F}\sigma}^2 - 2 k_{\Delta
\sigma}^2}$} & \mbox{for $k_{\mathrm{F}\sigma}^2 \ge 2 k_{\Delta
\sigma}^2$}\,\, , \\[0.2cm]
\left| \mu - \nu_\sigma \right|\,\, , & \mbox{occurring at
$\vek{k} = \vek{0}$} & \mbox{for $k_{\mathrm{F}\sigma}^2 < 2
k_{\Delta\sigma}^2$}\,\, .
\end{array}\right.
\end{equation}
Thus the criterion $\min |E_{\vek{k}\downarrow \eta}| = 0$ for
the topological transition~\cite{Sau2010,Alicea2010,Sato2010}
yields the condition
\begin{equation}
\mu = h_\mathrm{c} - \frac{|\Delta|^2}{2 h_\mathrm{c}}
\end{equation}
for the critical Zeeman energy within the approximate projected
theory, consistent with the leading-order small-$(|\Delta|/h)$
expansion of the exact relation $\mu = \sqrt{h_\mathrm{c}^2 -
|\Delta|^2}$ obtained by rearranging (\ref{eq:hcrit}) under the
assumption that $\mu > 0$. Furthermore, in the NSF phase, the
relations $0 < k_{\mathrm{F}\downarrow} < k_{\mathrm{F}
\uparrow}$ imply $|\bar{\Delta}_\downarrow| <
|\bar{\Delta}_\uparrow|$. Hence, the minimum
quasiparticle-excitation energy of the NSF is
$E_\mathrm{qp}^{(\mathrm{NSF})}\equiv \big|
\bar{\Delta}_\downarrow^{(\mathrm{NSF})}\big|$. This is in
contrast to the TSF phase where, except close to the transition,
the majority-spin excitation gap $|\bar{\Delta}_\uparrow|$ is
smaller than the minimum of the spin-$\downarrow$ dispersion at
$\vek{k}=\vek{0}$ and, thus, $E_\mathrm{qp}^{(\mathrm{TSF})}
\equiv \big|\bar{\Delta}_\uparrow^{(\mathrm{TSF})}\big|$. The
respective magnitudes of $E_\mathrm{qp}^{(\mathrm{TSF})}$ and
$E_\mathrm{qp}^{(\mathrm{NSF})}$ in a TSF-NSF hybrid system
generally depend on physical details. For our present work, we
envision a situation where $E_\mathrm{qp}^{(\mathrm{TSF})} <
E_\mathrm{qp}^{(\mathrm{NSF})}$, as illustrated in
Fig.~\ref{fig:ISSprime}(c).

Within our approach, the spin-labelled large components of each
Nambu four-spinor are associated with the quasiparticle dynamics
via the respective $2\times 2$ BdG equations (\ref{eq:2x2BdGup})
and (\ref{eq:2x2BdGdown}). However, the small components in the
four-spinors are not discarded but, rather, determined via the
relations (\ref{eq:2x2downUp}) and (\ref{eq:2x2upDown}) for
inclusion in relevant fomulae for physical quantities. This
formalism has been shown~\cite{Brand2018} to yield accurate
results for thermodynamic properties such as the
quasiparticle-density distribution and self-consistently
determined chemical and pair potentials. Here we apply this
method to the calculation of Andreev bound states in the
ISS$^\prime$ system depicted in Fig.~\ref{fig:ISSprime} for
which consideration of the entire four-spinor wave function is
essential~\cite{Brydon2015}. The formalism developed in the
remainder of this subsection forms the basis for the theoretical
description of the IS and SS$^\prime$ interfaces discussed in
Secs.~\ref{sec:ISform} and \ref{sec:SSPform} below.

\subsubsection{Wave numbers for interface matching}

We first obtain accurate approximations for the wave numbers
$k_\zeta^{(R)}$ appearing in the \textit{Ansatz\/}
(\ref{eq:genPWans}) by inverting the relations $E_{\vek{k}\sigma
\eta} = E$ for fixed $E$, using the quasiparticle-energy
dispersions from Eq.~(\ref{eq:lowEdisp}) within a particular
region $R\in \{ \mathrm{I}, \mathrm{S}, \mathrm{S}^\prime\}$.
This procedure yields
\begin{equation}\label{eq:ProjWaveVecs}
k^{(R)}_{\sigma\tau\alpha} = \alpha\, \sqrt{\left(
k_{\mathrm{F}\sigma}^{(R)}\right)^2 - k_y^2 - 2 \left( k_{\Delta
\sigma}^{(R)}\right)^2 +\tau\, \frac{2 m_\sigma^{(R)}}{\hbar^2}
\, \sqrt{E^2 - \left(\bar{\Delta}_\sigma^{(R)}\right)^2}}\quad ,
\end{equation}
where the superscript $(R)$ is now used liberally to indicate
that quantities pertain to region $R$. The label $\alpha \in\{ +
, - \}$ distinguishes right-moving ($\alpha=+$) and left-moving
($\alpha=-$) states, while $\tau\in\{+ , -\}$ labels
quasiparticle ($\tau=+$) and quasihole ($\tau=-$) excitations.
Figure~\ref{fig:WaveVecs} illustrates the origin of this
nomenclature, including the spin-$\uparrow$ (spin-$\downarrow$)
label for those wave numbers that arise from parts of the
dispersion approximated by the blue (red) dashed curves. The
Nambu two-spinor solutions of the spin-$\sigma$ $2\times 2$ BdG
equations [(\ref{eq:2x2BdGup}) for spin-$\uparrow$ and
(\ref{eq:2x2BdGdown}) for spin-$\downarrow$] associated with
energy $E$ and 2D wave vector $\kk^{(R)}_{\sigma\tau\alpha}=
\big( k^{(R)}_{\sigma\tau\alpha}, k_y\big)$ are found as
\begin{subequations}\label{eq:exactUV}
\begin{align}
u_{\sigma\tau\alpha}^{(R)} &= -\varsigma\, i\, \exp\left( i 
\varphi^{(R)} \right)\,\, \frac{k^{(R)}_{\sigma\tau\alpha} -
\varsigma \, i \, k_y}{\sqrt{\left( k^{(R)}_{\sigma\tau\alpha}
\right)^2 + k_y^2}} \, \sqrt{\frac{E -
\frac{\hbar^2}{m^{(R)}_\sigma}\, \left(k_{\Delta\sigma}^{(R)}
\right)^2 + \tau \, \sqrt{E^2 - \left( \bar{\Delta}^{(R)}_\sigma
\right)^2}}{2 E}} \quad , \\[0.2cm]
v_{\sigma\tau\alpha}^{(R)} &= \sgn(E) \,\, \sqrt{\frac{E +
\frac{\hbar^2}{m^{(R)}_\sigma}\, \left( k^{(R)}_{\Delta\sigma}
\right)^2 - \tau \, \sqrt{E^2 - \left( \bar{\Delta}^{(R)}_\sigma
\right)^2}}{2 E}} \quad ,
\end{align}
\end{subequations}
with $\sgn(\cdot)$ denoting the sign function and $\varsigma =
+$ ($-$) for $\sigma=\,\,\uparrow$ ($\downarrow$). Using also
the relations (\ref{eq:2x2downUp}) and (\ref{eq:2x2upDown}) for
a Nambu four-spinor's small components in terms of the
dominant-spin ones, it is possible to express the
Feshbach-projection-approximated form of the full four-component
Nambu eigenspinor of the BdG equation (\ref{eq:4x4ham}) in
region $R$ with energy $E$ and wave vector $\kk^{(R)}_{\sigma
\tau\alpha}$ in terms of that eigenspinor's dominating
(large-spin-$\sigma$) part $\big(u_{\sigma\tau\alpha}^{(R)},
v_{\sigma\tau\alpha}^{(R)}\big)^T$ as 
\begin{equation}
\begin{pmatrix} u_{\uparrow\kk_{\sigma\tau\alpha}^{(R)}}
\\[0.1cm] v_{\uparrow\kk_{\sigma\tau\alpha}^{(R)}} \\[0.1cm]
u_{\downarrow\kk_{\sigma\tau\alpha}^{(R)}}\\[0.1cm]
v_{\downarrow\kk_{\sigma\tau\alpha}^{(R)}} \end{pmatrix} =
\mathcal{M}_{\sigma\tau\alpha}^{(R)}\,\,\, \begin{pmatrix}
u_{\sigma\tau\alpha}^{(R)} \\[5pt] v_{\sigma\tau\alpha}^{(R)}
\end{pmatrix} \quad .
\end{equation}
The $4\times 2$ matrices
\begin{equation}
\mathcal{M}_{\uparrow\tau\alpha}^{(R)} = \begin{pmatrix}
\mathbbm{1}_{2\times 2} \\[0.2cm] \mathcal{J}_{\uparrow\tau
\alpha}^{(R)} \end{pmatrix}\quad , \quad
\mathcal{M}_{\downarrow\tau\alpha}^{(R)} = \begin{pmatrix}
\mathcal{J}_{\downarrow\tau\alpha}^{(R)} \\[0.2cm] 
\mathbbm{1}_{2\times 2} \end{pmatrix}
\end{equation}
contain the $2\times 2$ unit matrix $\mathbbm{1}_{2\times 2}$,
and the remaining $2\times 2$ blocks
\begin{equation}\label{eqs:MandJ}
\mathcal{J}_{\uparrow\tau\alpha}^{(R)} = \frac{1}{2 h}
\begin{pmatrix} \lambda \big( i k_{\uparrow\tau\alpha}^{(R)} -
\, k_y \big) & -\Delta^{(R)} \\[0.2cm] -\big( \Delta^{(R)}
\big)^\ast & \lambda \big( i k_{\uparrow\tau\alpha}^{(R)} +\,
k_y \big) \end{pmatrix} \,\,\, , \,\,\,
\mathcal{J}_{\downarrow\tau\alpha}^{(R)} = \frac{1}{2h}
\begin{pmatrix} \lambda \big( i k_{\downarrow\tau\alpha}^{(R)} +
\, k_y \big) & -\Delta^{(R)} \\[0.2cm] -\big( \Delta^{(R)}
\big)^\ast & \lambda \big( i k_{\downarrow\tau\alpha}^{(R)} -\,
k_y \big) \end{pmatrix}
\end{equation}
are obtained by specializing the matrices entering the relations
(\ref{eq:2x2downUp}) and (\ref{eq:2x2upDown}) between small and
large components of Nambu four-spinors to the situation where
$\kk=\kk^{(R)}_{\sigma\tau\alpha}$.

\subsubsection{Andreev approximation}

In the following two subsections \ref{sec:ISform} and
\ref{sec:SSPform}, the general formalism developed here will be
applied to describe bound states at the IS and $\mathrm{S}
\mathrm{S}^\prime$ interfaces, respectively. Useful physical
insights emerge from analytical results that are obtained using
the Andreev approximation~\cite{Andreev1964}. In the present
context, this approach amounts to neglecting $k_{\Delta\sigma}$
and further  approximating Eq.~\eqref{eq:ProjWaveVecs} by
\begin{equation}\label{eq:kAA}
k^{(R)}_{\sigma\tau\alpha} \approx \alpha\, k_{\mathrm{F}
\sigma}^{(R)} + \alpha\tau\, \frac{m_\sigma^{(R)}}{\hbar^2
k_{\mathrm{F}\sigma}^{(R)}}\, \sqrt{E^2 - \left(
\Delta^{(R)}_\sigma \right)^2} \quad .
\end{equation}
To ensure that (\ref{eq:kAA}) is a good approximation, $|E|,
|\Delta^{(R)}_\sigma| \ll |\mu - \nu_\sigma^{(R)}|$ and $|k_y|
\ll |k^{(R)}_{\mathrm{F}\sigma}|$ are required, which also
guarantee $|k_{\Delta\sigma}| \ll |k^{(R)}_{\mathrm{F}\sigma}|$.
More explicitly, the condition for the validity of the Andreev
approximation is expressed as
\begin{equation}\label{eq:AAexplicit}
1\gg \frac{m_\sigma^{(R)}\, \big|\Delta^{(R)}_\sigma
\big|}{\hbar^2\big|k_{\mathrm{F}\sigma}^{(R)}\big|^2} \equiv
\frac{\lambda\, \big|k_{\mathrm{F}\sigma}^{(R)}\big|}{h}\,
\frac{m_\sigma^{(R)}\, \big|\Delta^{(R)}\big|}{\hbar^2 \big|
k_{\mathrm{F}\sigma}^{(R)}\big|^2} = \mathcal{O}\left(
\sqrt{\frac{m\, \lambda^2}{\hbar^2 h}} \right)\cdot \mathcal{O}
\left( \sqrt{\frac{m_\sigma^{(R)}\, \big|\Delta^{(R)}
\big|^2}{\hbar^2\big| k_{\mathrm{F}\sigma}^{(R)}\big|^2 h}}
\right) \quad .
\end{equation}
Thus, in our system of interest, applicability of the Andreev
approximation can be guaranteed by the smallness of the
\textit{s}-wave pair potential compared to the Fermi energy (the
usual condition~\cite{Andreev1964}) and/or a small magnitude of
the spin-orbit coupling. Using the Andreev approximation in the
Bogoliubov-quasiparticle spinors yields
\begin{subequations}\label{eq:uvAA}
\begin{align}\label{eq:uAA}
u_{\sigma\tau\alpha}^{(R)} &\approx \left\{ \begin{array}{cl}
-\varsigma \alpha\, i\, \exp\left[ i \left( \varphi^{(R)} +
\frac{\tau\, \sgn(E)}{2}\, \theta^{(R)}_\sigma - \varsigma
\alpha \, \vartheta_{k_y\sigma}^{(R)} \right) \right]\,\,
\sqrt{\frac{\Delta^{(R)}_\sigma}{2|E|}}\,\, , & k_{\mathrm{F}
\sigma}^{(R)} = \big|k_{\mathrm{F}\sigma}^{(R)}\big| \\[0.3cm]
-\varsigma \alpha\, i\, \exp\left( i \varphi^{(R)} \right)\,\,
\sqrt{\frac{|k_{\mathrm{F}\sigma}^{(R)}| - \varsigma \alpha\,
k_y}{|k_{\mathrm{F}\sigma}^{(R)}| + \varsigma \alpha\,k_y}}\,\,
\,\sqrt{\frac{E + \tau \, \sqrt{E^2 + \left|\Delta^{(R)}_\sigma
\right|^2}}{2 E}}\,\, , & k_{\mathrm{F}\sigma}^{(R)} = i\,
\big|k_{\mathrm{F}\sigma}^{(R)}\big| \end{array}\right. ,
\\[0.5cm] v_{\sigma\tau\alpha}^{(R)} &\approx \left\{
\begin{array}{cl} \sgn(E)\,\, \exp\left( - i \, \frac{\tau \,
\sgn(E)}{2} \, \theta_\sigma^{(R)} \right)\,\,
\sqrt{\frac{\Delta^{(R)}_\sigma}{2|E|}} \,\, , & k_{\mathrm{F}
\sigma}^{(R)} = \big|k_{\mathrm{F}\sigma}^{(R)}\big| \\[0.3cm]
\sgn(E)\,\, \sqrt{\frac{E - \tau \, \sqrt{E^2 +
\left|\Delta^{(R)}_\sigma\right|^2}}{2 E}}\,\, , & k_{\mathrm{F}
\sigma}^{(R)} = i\, \big|k_{\mathrm{F}\sigma}^{(R)}\big|
\end{array} \right. ,
\end{align}
\end{subequations}
with $\theta^{(R)}_\sigma = \arccos\big(|E|/\Delta^{(R)}_\sigma
\big)$, $\vartheta^{(R)}_{k_y\sigma} = \arcsin\big(k_y/
k^{(R)}_{\mathrm{F}\sigma}\big)$, and again $\varsigma = +$
($-$) for $\sigma=\,\uparrow$ ($\downarrow$).

\subsection{IS interface --- edge states}
\label{sec:ISform}

Our particular system of interest has an IS interface where the
S region is in the TSF phase [see Fig.~\ref{fig:ISSprime}(a)].
For completeness and to provide further insight through
comparisons, we consider in the following also the case of an IS
interface where S is an NSF.

We assume that the I region is inaccessible to quasiparticle
excitations, i.e., its presence imposes the hard-wall boundary
condition
\begin{equation}\label{eq:Ibc}
\begin{pmatrix}
u_{\uparrow}^{(\mathrm{S})}(-L) \\[0.1cm]
v_{\uparrow}^{(\mathrm{S})}(-L) \\[0.1cm]
u_{\downarrow}^{(\mathrm{S})}(-L) \\[0.1cm]
v_{\downarrow}^{(\mathrm{S})}(-L) \end{pmatrix} =
\begin{pmatrix} 0 \\[0.2cm] 0 \\[0.2cm] 0 \\[0.2cm] 0
\end{pmatrix} \quad .
\end{equation}
To describe Andreev bound states localized at the IS interface
via the \textit{Ansatz}\/ (\ref{eq:genPWans}) with $\kk_\zeta
\to \kk^{(S)}_{\sigma\tau\alpha} = \big( k^{(S)}_{\sigma\tau
\alpha}, k_y\big)$, only evanescent states having $\Im\big(
k^{(\mathrm{S})}_{\sigma\tau\alpha}\big)>0$ should be included.
Inspection of Eqs.~(\ref{eq:ProjWaveVecs}) and/or (\ref{eq:kAA})
shows that, in regions where $\mu-\nu_\sigma^{(R)}>0$ and
$k^{(R)}_{\mathrm{F}\sigma}$ is therefore real, $k^{(R)}_{\sigma
\tau\alpha}$ becomes complex-valued for $|E| <
\Delta^{(R)}_\sigma$, i.e., subgap-energy states. If $\mu-
\nu_\sigma^{(R)}<0$ in region $R$, $k^{(R)}_{\mathrm{F}\sigma}$
and $k^{(R)}_{\sigma\tau\alpha}$ are both purely imaginary for
subgap-energy states. Hence, when S is in the NSF phase, states
to include in the \textit{Ansatz}\/ (\ref{eq:genPWans}) satisfy
$|E| < E_\mathrm{qp}^{(\mathrm{NSF})}\approx
\Delta_\downarrow^{(\mathrm{S})}$ and $\alpha\tau = +$. In
contrast, when the S region hosts a TSF, $|E| <
E_\mathrm{qp}^{(\mathrm{TSF})}\approx
\Delta_\uparrow^{(\mathrm{S})}$ is required together with
$\alpha\tau = +$ for $\sigma = \,\,\uparrow$ and $\alpha = +$
for $\sigma = \,\,\downarrow$. Based on these considerations, we
find the explicit form of the Nambu spinor for S being a NSF as
\begin{equation}\label{eq:I-NSF}
\begin{pmatrix}
u_{\uparrow}^{(\mathrm{NSF})}(x) \\[0.1cm]
v_{\uparrow}^{(\mathrm{NSF})}(x) \\[0.1cm]
u_{\downarrow}^{(\mathrm{NSF})}(x) \\[0.1cm]
v_{\downarrow}^{(\mathrm{NSF})}(x) \end{pmatrix} =  \sum_\sigma
\left[ a_{\sigma ++}^{(\mathrm{S})} \,\,\, \mathcal{M}_{\sigma
++}^{(\mathrm{S})}\,\,\, \begin{pmatrix} u_{\sigma
++}^{(\mathrm{S})} \\[5pt] v_{\sigma++}^{(\mathrm{S})}
\end{pmatrix} \,\,\, \ee^{i\, k_{\sigma ++}^{(\mathrm{S})}\, x}
+ a_{\sigma --}^{(\mathrm{S})} \,\,\, \mathcal{M}_{\sigma
--}^{(\mathrm{S})}\,\,\, \begin{pmatrix} u_{\sigma
--}^{(\mathrm{S})} \\[5pt] v_{\sigma --}^{(S)} \end{pmatrix}
\,\,\, \ee^{i\, k_{\sigma --}^{(\mathrm{S})}\, x} \right]\,\, ,
\end{equation}
while the allowed superposition for the case where S is a TSF
has the form
\begin{align}\label{eq:I-TSF}
\begin{pmatrix}
u_{\uparrow}^{(\mathrm{TSF})}(x) \\[0.1cm]
v_{\uparrow}^{(\mathrm{TSF})}(x) \\[0.1cm]
u_{\downarrow}^{(\mathrm{TSF})}(x) \\[0.1cm]
v_{\downarrow}^{(\mathrm{TSF})}(x)\end{pmatrix} &=  a_{\uparrow
++}^{(\mathrm{S})} \,\,\, \mathcal{M}_{\uparrow
++}^{(\mathrm{S})}\,\,\, \begin{pmatrix} u_{\uparrow
++}^{(\mathrm{S})} \\[5pt] v_{\uparrow ++}^{(\mathrm{S})}
\end{pmatrix}\,\,\, \ee^{i\, k_{\uparrow ++}^{(\mathrm{S})}\, x}
+ a_{\uparrow --}^{(\mathrm{S})}\,\,\, \mathcal{M}_{\uparrow
--}^{(\mathrm{S})}\,\,\, \begin{pmatrix} u_{\uparrow
--}^{(\mathrm{S})} \\[5pt] v_{\uparrow --}^{(\mathrm{S})}
\end{pmatrix} \,\,\, \ee^{i\, k_{\uparrow --}^{(\mathrm{S})}\,
x} \nonumber \\ &\hspace{0.3cm} +  a_{\downarrow
++}^{(\mathrm{S})} \,\,\, \mathcal{M}_{\downarrow
++}^{(\mathrm{S})}\,\,\, \begin{pmatrix} u_{\downarrow
++}^{(\mathrm{S})} \\[5pt] v_{\downarrow ++}^{(\mathrm{S})}
\end{pmatrix}\,\,\, \ee^{i\, k_{\downarrow ++}^{(\mathrm{S})}\,
x} + a_{\downarrow -+}^{(\mathrm{S})}\,\,\,
\mathcal{M}_{\downarrow -+}^{(\mathrm{S})}\,\,\, \begin{pmatrix}
u_{\downarrow -+}^{(\mathrm{S})} \\[5pt] v_{\downarrow
-+}^{(\mathrm{S})} \end{pmatrix} \,\,\, \ee^{i\, k_{\downarrow
-+}^{(\mathrm{S})}\, x} \,\,\, .
\end{align}
Requiring (\ref{eq:Ibc}) to hold for \textit{Ansatz}\/
(\ref{eq:I-NSF}) or (\ref{eq:I-TSF}), respectively, yields a
homogeneous system of four linear equations for the four
coefficients $a_{\sigma\tau\alpha}^{(\mathrm{S})}$ appearing in
that particular \textit{Ansatz\/} whose characteristic equation
determines the Andreev-edge-state energies. As the fully
general form of the characteristic equations is quite long and
rather unilluminating~\cite{Thompson2022}, we only present
results here after applying further approximations.

If entries $\mathcal{O}\big(|\Delta^{(R)}|/h\big)$ in the
matrices $\mathcal{J}_{\sigma\tau\alpha}^{(R)}$ defined in
Eq.~(\ref{eqs:MandJ}) are neglected, we obtain the result
\begin{align} \label{eq:NSFsec}
&\left\{\left[ 1 - \frac{\lambda^2}{4 h^2}\, i\, k_y \left(
k_{\uparrow ++}^{(\mathrm{S})} - k_{\uparrow --}^{(\mathrm{S})}
\right) \right] u_{\uparrow ++}^{(\mathrm{S})}\,
v_{\uparrow --}^{(\mathrm{S})} - \left[ 1 + \frac{\lambda^2}{4
h^2}\, i\, k_y \left( k_{\uparrow ++}^{(\mathrm{S})} -
k_{\uparrow --}^{(\mathrm{S})} \right) \right] u_{\uparrow
--}^{(\mathrm{S})}\, v_{\uparrow ++}^{(\mathrm{S})}\right\}
\nonumber \\
& \times \left\{\left[ 1 + \frac{\lambda^2}{4 h^2}\, i\, k_y
\left( k_{\downarrow ++}^{(\mathrm{S})} - k_{\downarrow 
--}^{(\mathrm{S})} \right) \right] u_{\downarrow
++}^{(\mathrm{S})}\, v_{\downarrow --}^{(\mathrm{S})} - \left[
1 - \frac{\lambda^2}{4 h^2}\, i\, k_y \left( k_{\downarrow 
++}^{(\mathrm{S})} - k_{\downarrow --}^{(\mathrm{S})} \right)
\right] u_{\downarrow --}^{(\mathrm{S})}\, v_{\downarrow
++}^{(\mathrm{S})}\right\} \nonumber \\
&= \frac{\lambda^2}{4 h^2} \left( k_{\uparrow ++}^{(\mathrm{S})}
- k_{\uparrow --}^{(\mathrm{S})} \right) \left( k_{\downarrow 
++}^{(\mathrm{S})} - k_{\downarrow --}^{(\mathrm{S})} \right)
\left( u_{\uparrow ++}^{(\mathrm{S})}\, v_{\downarrow
++}^{(\mathrm{S})} - u_{\downarrow ++}^{(\mathrm{S})}\,
v_{\uparrow ++}^{(\mathrm{S})} \right) \left( u_{\uparrow
--}^{(\mathrm{S})}\, v_{\downarrow --}^{(\mathrm{S})} -
u_{\downarrow --}^{(\mathrm{S})}\, v_{\uparrow
--}^{(\mathrm{S})} \right)
\end{align}
for the characteristic equation in the NSF case. As the
\textit{Ans\"atze\/} Eqs.~(\ref{eq:I-NSF}) and (\ref{eq:I-TSF})
for the I-NSF and I-TSF boundaries differ only by the
replacement of the state with label $\sigma\tau\alpha \equiv\,\,
\downarrow\! -\, -$ in the former by the state labelled $\sigma
\tau\alpha \equiv \,\, \downarrow\! -\, +$ in the latter, the
characteristic equation for the situation where S is in the TSF
phase can be obtained by making the corresponding adjustments in
Eq.~(\ref{eq:NSFsec}), yielding
\begin{align} \label{eq:TSFsec}
&\left\{ \left[ 1 - \frac{\lambda^2}{4 h^2}\, i\, k_y \left(
k_{\uparrow ++}^{(\mathrm{S})} - k_{\uparrow --}^{(\mathrm{S})}
\right) \right] u_{\uparrow ++}^{(\mathrm{S})}\,
v_{\uparrow --}^{(\mathrm{S})} - \left[ 1 + \frac{\lambda^2}{4
h^2}\, i\, k_y \left( k_{\uparrow ++}^{(\mathrm{S})} -
k_{\uparrow --}^{(\mathrm{S})} \right) \right] u_{\uparrow
--}^{(\mathrm{S})}\, v_{\uparrow ++}^{(\mathrm{S})} \right\}
\nonumber \\
& \times \left\{ \left[ 1 + \frac{\lambda^2}{4 h^2}\, i\, k_y
\left( k_{\downarrow ++}^{(\mathrm{S})} - k_{\downarrow 
-+}^{(\mathrm{S})} \right) \right] u_{\downarrow
++}^{(\mathrm{S})}\, v_{\downarrow -+}^{(\mathrm{S})} - \left[
1 - \frac{\lambda^2}{4 h^2}\, i\, k_y \left( k_{\downarrow 
++}^{(\mathrm{S})} - k_{\downarrow -+}^{(\mathrm{S})} \right)
\right] u_{\downarrow -+}^{(\mathrm{S})}\, v_{\downarrow
++}^{(\mathrm{S})} \right\} \nonumber \\
&= \frac{\lambda^2}{4 h^2} \left( k_{\uparrow ++}^{(\mathrm{S})}
- k_{\uparrow --}^{(\mathrm{S})} \right) \left( k_{\downarrow 
++}^{(\mathrm{S})} - k_{\downarrow -+}^{(\mathrm{S})} \right)
\left( u_{\uparrow ++}^{(\mathrm{S})}\, v_{\downarrow
++}^{(\mathrm{S})} - u_{\downarrow ++}^{(\mathrm{S})}\,
v_{\uparrow ++}^{(\mathrm{S})} \right) \left( u_{\uparrow
--}^{(\mathrm{S})}\, v_{\downarrow -+}^{(\mathrm{S})} -
u_{\downarrow -+}^{(\mathrm{S})}\, v_{\uparrow
--}^{(\mathrm{S})} \right) \,\, .
\end{align}
Application of the Andreev-approximation expressions from
Eqs.~(\ref{eq:kAA}) and (\ref{eq:uvAA}) transforms the
characteristic equation (\ref{eq:NSFsec}) for the I-NSF system
into the form
\begin{subequations}
\begin{align}
& \cos\left[ \theta^{(\mathrm{S})}_\uparrow - \sgn(E) \left(
\vartheta^{(\mathrm{S})}_{k_y\uparrow} + \frac{\lambda^2}{2 h^2}
\, k^{(\mathrm{S})}_{\mathrm{F}\uparrow}\, k_y\right) \right]
\,\, \cos\left[\theta^{(\mathrm{S})}_\downarrow + \sgn(E)\,
\left( \vartheta^{(\mathrm{S})}_{k_y\downarrow} +
\frac{\lambda^2}{2 h^2}\, k^{(\mathrm{S})}_{\mathrm{F}
\downarrow}\, k_y\right) \right] \nonumber \\[0.2cm] &
\hspace{1cm} = \frac{\lambda^2 k^{(\mathrm{S})}_{\mathrm{F}
\uparrow} k^{(\mathrm{S})}_{\mathrm{F}\downarrow}}{h^2}\,\,
\cos^2 \left[ \frac{1}{2}\left( \theta^{(\mathrm{S})}_\uparrow -
\sgn(E)\,\, \vartheta^{(\mathrm{S})}_{k_y\uparrow} \right) -
\frac{1}{2}\left(\theta^{(\mathrm{S})}_\downarrow + \sgn(E)\,\,
\vartheta^{(\mathrm{S})}_{k_y\downarrow} \right)\right]
\\[0.2cm] \label{eq:NSFsecAA} & \hspace{1cm} \approx
\frac{\lambda^2 k^{(\mathrm{S})}_{\mathrm{F}\uparrow}
k^{(\mathrm{S})}_{\mathrm{F}\downarrow}}{2 h^2} \left\{ 1 +
\sin\left[ \theta^{(\mathrm{S})}_\uparrow - \sgn(E)\, \left(
\vartheta^{(\mathrm{S})}_{k_y\uparrow} + \frac{\lambda^2}{2
h^2}\, k^{(\mathrm{S})}_{\mathrm{F}\uparrow}\, k_y\right)\right]
\right.  \nonumber \\[0.2cm] & \hspace{6cm} \left. \times
\sin\left[ \theta^{(\mathrm{S})}_\downarrow + \sgn(E)\, \left(
\vartheta^{(\mathrm{S})}_{k_y\downarrow} + \frac{\lambda^2}{2
h^2}\, k^{(\mathrm{S})}_{\mathrm{F}\downarrow}\, k_y\right)
\right] \right\} \,\, ,
\end{align}
\end{subequations}
where obtaining the right-hand side of (\ref{eq:NSFsecAA})
involves omitting higher-order corrections of the type we have
neglected all along. Similarly, the Andreev-approximated form of
the characteristic equation (\ref{eq:TSFsec}) for the I-TSF
boundary becomes
\begin{align}\label{eq:TSFsecAA}
\cos\left[ \theta^{(\mathrm{S})}_\uparrow - \sgn(E)\, \left(
\vartheta^{(\mathrm{S})}_{k_y\uparrow} + \frac{\lambda^2}{2
h^2}\, k^{(\mathrm{S})}_{\mathrm{F}\uparrow}\, k_y \right)
\right] = \mathcal{O}\left( \frac{\lambda^2 }{h^2}\,
k^{(\mathrm{S})}_{\mathrm{F}\uparrow}\,
|k^{(\mathrm{S})}_{\mathrm{F}\downarrow}|\,
\frac{m^{(\mathrm{S})}_\downarrow E}{\hbar^2 |\mu^{(\mathrm{S})}
- \nu^{(\mathrm{S})}_\downarrow|} \right) \approx 0 \quad .
\end{align}
We present solutions of Eqs.~(\ref{eq:NSFsecAA}) and
(\ref{eq:TSFsecAA}), and discuss their physical meaning, in
Sec.~\ref{sec:resSurf}.

Before concluding this subsection, we present an alternative
approach to treating the TSF edge that is instructive for our
later consideration of the TSF-NSF interface. The system of
linear equations arising from requiring the boundary condition
(\ref{eq:Ibc}) for the \textit{Ansatz}\/ (\ref{eq:I-TSF}) can
be written as
\begin{equation}\label{eq:TSFalg}
\begin{pmatrix} \mathcal{D}_\uparrow^{(\mathrm{S})} &
\mathcal{C}_{\uparrow\downarrow}^{(\mathrm{S})} \\[0.5cm]
\mathcal{C}_{\downarrow\uparrow}^{(\mathrm{S})} &
\mathcal{D}_\downarrow^{(\mathrm{S})} \end{pmatrix}
\begin{pmatrix} a_{\uparrow ++}^{(\mathrm{S})} \\[0.1cm]
a_{\uparrow --}^{(\mathrm{S})} \\[0.1cm] a_{\downarrow
++}^{(\mathrm{S})} \\[0.1cm] a_{\downarrow -+}^{(\mathrm{S})}
\end{pmatrix} = \begin{pmatrix} 0 \\[0.2cm] 0 \\[0.2cm] 0
\\[0.2cm] 0 \end{pmatrix} \quad ,
\end{equation}
with the $2\times 2$-matrix entries
\begin{subequations}
\begin{align}
& \mathcal{D}_\uparrow^{(\mathrm{S})} = \begin{pmatrix}
u_{\uparrow ++}^{(\mathrm{S})} & u_{\uparrow --}^{(\mathrm{S})}
\\[0.2cm] v_{\uparrow ++}^{(\mathrm{S})} & v_{\uparrow
--}^{(\mathrm{S})} \end{pmatrix} \quad , \quad
\mathcal{C}_{\downarrow\uparrow}^{(\mathrm{S})} =
\begin{pmatrix} \mathcal{J}_{\uparrow ++}^{(\mathrm{S})}
\begin{pmatrix} u_{\uparrow ++}^{(\mathrm{S})} \\[0.1cm]
v_{\uparrow ++}^{(\mathrm{S})}\end{pmatrix} &
\mathcal{J}_{\uparrow --}^{(\mathrm{S})} \begin{pmatrix}
u_{\uparrow --}^{(\mathrm{S})} \\[0.1cm] v_{\uparrow
--}^{(\mathrm{S})} \end{pmatrix} \end{pmatrix} \quad , \\[0.1cm]
& \mathcal{D}_\downarrow^{(\mathrm{S})} = \begin{pmatrix}
u_{\downarrow ++}^{(\mathrm{S})} & u_{\downarrow
-+}^{(\mathrm{S})}\\[0.2cm] v_{\downarrow ++}^{(\mathrm{S})} &
v_{\downarrow -+}^{(\mathrm{S})} \end{pmatrix} \quad , \quad
\mathcal{C}_{\uparrow\downarrow}^{(\mathrm{S})} =
\begin{pmatrix} \mathcal{J}_{\downarrow ++}^{(\mathrm{S})}
\begin{pmatrix} u_{\downarrow ++}^{(\mathrm{S})} \\[0.1cm]
v_{\downarrow ++}^{(\mathrm{S})}\end{pmatrix} & 
\mathcal{J}_{\downarrow -+}^{(\mathrm{S})} \begin{pmatrix}
u_{\downarrow -+}^{(\mathrm{S})} \\[0.1cm] v_{\downarrow
-+}^{(\mathrm{S})} \end{pmatrix} \end{pmatrix} \quad .
\end{align}
\end{subequations}
The condition $\mathrm{det}\big(
\mathcal{D}_\uparrow^{(\mathrm{S})}\big)=0$ would yield the
familiar~\cite{Honerkamp1998} chiral-\textit{p}-wave Majorana
edge mode for the isolated spin-$\uparrow$ sector, but the
coupling between the spin-$\uparrow$ and spin-$\downarrow$
sectors embodied in the matrices $\mathcal{C}_{\uparrow
\downarrow}^{(\mathrm{S})}$ and $\mathcal{C}_{\downarrow
\uparrow}^{(\mathrm{S})}$ leads to modifications.
Straightforward elimination of the coefficients $a_{\downarrow
++}^{(\mathrm{S})}$ and $a_{\downarrow -+}^{(\mathrm{S})}$ in
Eq.~(\ref{eq:TSFalg}) yields the equivalent set
\begin{equation}\label{eq:sUpEquEdge}
\begin{pmatrix} a_{\downarrow ++}^{(\mathrm{S})} \\[0.1cm]
a_{\downarrow -+}^{(\mathrm{S})} \end{pmatrix} = -\left[
\mathcal{D}_\downarrow^{(\mathrm{S})} \right]^{-1}
\mathcal{C}_{\downarrow\uparrow}^{(\mathrm{S})} \,\,
\begin{pmatrix} a_{\uparrow ++}^{(\mathrm{S})} \\[0.1cm]
a_{\uparrow --}^{(\mathrm{S})} \end{pmatrix} \quad , \quad
\left\{ \mathcal{D}_\uparrow^{(\mathrm{S})} -
\mathcal{C}_{\uparrow\downarrow}^{(\mathrm{S})} \left[
\mathcal{D}_\downarrow^{(\mathrm{S})} \right]^{-1}
\mathcal{C}_{\downarrow\uparrow}^{(\mathrm{S})} \right\}
\begin{pmatrix} a_{\uparrow ++}^{(\mathrm{S})} \\[0.1cm]
a_{\uparrow --}^{(\mathrm{S})} \end{pmatrix}  = \begin{pmatrix}
0 \\[0.2cm] 0 \end{pmatrix}
\end{equation}
of linear equations, and the characteristic equation for the
Andreev-edge-state energy becomes
\begin{equation}\label{eq:TSFsec2x2}
\mathrm{det} \left( \mathcal{D}_\uparrow^{(\mathrm{S})} -
\mathcal{C}_{\uparrow\downarrow}^{(\mathrm{S})} \left[
\mathcal{D}_\downarrow^{(\mathrm{S})} \right]^{-1}
\mathcal{C}_{\downarrow\uparrow}^{(\mathrm{S})}\right) = 0
\quad .
\end{equation}
Considering the matrices depending on spin-$\downarrow$
Nambu-spinor amplitudes, we find
\begin{subequations}
\begin{align}
\mathcal{C}_{\uparrow\downarrow}^{(\mathrm{S})} \left[
\mathcal{D}_\downarrow^{(\mathrm{S})} \right]^{-1} &= \frac{1}{2
h} \begin{pmatrix} \lambda \left[ \frac{i}{2} \left(
k_{\downarrow ++}^{(\mathrm{S})} +
k_{\downarrow-+}^{(\mathrm{S})} \right) + \, k_y \right] &
-\Delta^{(\mathrm{S})} \\[0.2cm] -\big( \Delta^{(\mathrm{S})}
\big)^\ast & \lambda \left[ \frac{i}{2} \left( k_{\downarrow
++}^{(\mathrm{S})} + k_{\downarrow-+}^{(\mathrm{S})} \right) -
\, k_y \right] \end{pmatrix} \nonumber \\[0.2cm] &
\hspace{6.5cm} + \mathcal{O} \left( \frac{\lambda
k_{\mathrm{F}\downarrow}^{(\mathrm{S})}}{h} \,
\frac{m_\downarrow^{(\mathrm{S})} \big|
\Delta_\downarrow^{(\mathrm{S})}\big|}{\hbar^2 \big|
k_{\mathrm{F}\downarrow}^{(\mathrm{S})} \big|^2} \right) \quad ,
\\ &\approx \frac{1}{2 h} \begin{pmatrix} -\lambda \left( \big|
k_{\mathrm{F}\downarrow}^{(\mathrm{S})} \big| - k_y \right) & 0
\\[0.1cm] 0 & -\lambda \left( \big| k_{\mathrm{F}
\downarrow}^{(\mathrm{S})} \big| + k_y \right) \end{pmatrix}
\quad , \label{eq:spinDownMatS}
\end{align}
\end{subequations}
where the Andreev approximation has been employed in the steps
leading to Eq.~(\ref{eq:spinDownMatS}). Using the form
(\ref{eq:spinDownMatS}) in (\ref{eq:TSFsec2x2}) and retaining
only leading-order corrections in the spin-orbit-coupling
strength yields the previously obtained characteristic equation
(\ref{eq:TSFsecAA}). While ultimately giving the same result,
the alternative approach required only the manipulation of $2
\times 2$ matrices instead of considering the full $4\times
4$-matrix determinant arising from the condition
(\ref{eq:TSFalg}). Although either way was possible to be
pursued successfully in the case of the TSF edge, the analogous
dimensional reduction of the $8\times 8$ problem emerging for
the TSF-NSF interface (discussed in the following subsection
\ref{sec:SSPform}) to a more easily tractable $4 \times 4$
system of linear equations will prove crucial for obtaining
analytical results in that situation. Furthermore, writing the
characteristic equation in the form (\ref{eq:TSFsec2x2}) makes
the modifications arising from the coupling between
spin-$\uparrow$ and spin-$\downarrow$ sectors more explicit,
thus also enabling a more systematic approach to introducing
approximations.

\subsection{SS$^\prime$ interface --- Josephson junction}
\label{sec:SSPform}

Describing the SS$^\prime$ interface located at $x=0$ requires
matching superpositions of evanescent-quasiparticle excitations
from the two adjoining regions as per the conditions
\begin{equation}\label{eq:SSpConti}
\begin{pmatrix}
u_{\uparrow}^{(\mathrm{S})}(0) \\[0.1cm]
v_{\uparrow}^{(\mathrm{S})}(0) \\[0.1cm]
u_{\downarrow}^{(\mathrm{S})}(0) \\[0.1cm]
v_{\downarrow}^{(\mathrm{S})}(0)\end{pmatrix} =
\begin{pmatrix}
u_{\uparrow}^{(\mathrm{S}^\prime)}(0) \\[0.1cm]
v_{\uparrow}^{(\mathrm{S}^\prime)}(0) \\[0.1cm]
u_{\downarrow}^{(\mathrm{S}^\prime)}(0) \\[0.1cm]
v_{\downarrow}^{(\mathrm{S}^\prime)}(0)\end{pmatrix} \quad ,
\quad \left. \frac{d}{d x} \begin{pmatrix}
u_{\uparrow}^{(\mathrm{S})}(x) \\[0.1cm]
v_{\uparrow}^{(\mathrm{S})}(x) \\[0.1cm]
u_{\downarrow}^{(\mathrm{S})}(x) \\[0.1cm]
v_{\downarrow}^{(\mathrm{S})}(x)\end{pmatrix} \right|_{x=0} =
\left. \frac{d}{d x} \begin{pmatrix}
u_{\uparrow}^{(\mathrm{S}^\prime)}(x) \\[0.1cm]
v_{\uparrow}^{(\mathrm{S}^\prime)}(x) \\[0.1cm]
u_{\downarrow}^{(\mathrm{S}^\prime)}(x) \\[0.1cm]
v_{\downarrow}^{(\mathrm{S}^\prime)}(x)\end{pmatrix}
\right|_{x=0} .
\end{equation}
Here the general \textit{Ansatz}\/ (\ref{eq:genPWans}) for the S
[S$^\prime$] part occupying the half-space $x<0$ [$x>0$]
contains the four Nambu spinors with $\kk_\zeta =
\kk^{(S)}_{\sigma\tau\alpha} \equiv \big( k^{(S)}_{\sigma\tau
\alpha}, k_y\big)$ for which $\Im \big(k^{(\mathrm{S})}_{\sigma
\tau\alpha}\big) < 0$ $\big[\Im
\big(k^{(\mathrm{S}^\prime)}_{\sigma\tau\alpha}\big) > 0\big]$.
As we consider the situation where S$^\prime$ is a NSF, the form
of the superposition shown on the right-hand-side of
Eq.~(\ref{eq:I-NSF}) applies; except that S$^\prime$ should
replace S in all superscripts. The explicit form of the
\textit{Ansatz}\/ for the TSF in the S region reads
\begin{align}\label{eq:S_TSF}
\begin{pmatrix}
u_{\uparrow}^{(\mathrm{S})}(x) \\[0.1cm]
v_{\uparrow}^{(\mathrm{S})}(x) \\[0.1cm]
u_{\downarrow}^{(\mathrm{S})}(x) \\[0.1cm]
v_{\downarrow}^{(\mathrm{S})}(x)\end{pmatrix} &=  a_{\uparrow
+-}^{(\mathrm{S})} \,\,\, \mathcal{M}_{\uparrow
+-}^{(\mathrm{S})}\,\,\, \begin{pmatrix} u_{\uparrow
+-}^{(\mathrm{S})} \\[5pt] v_{\uparrow +-}^{(\mathrm{S})}
\end{pmatrix}\,\,\, \ee^{i\, k_{\uparrow +-}^{(\mathrm{S})}\, x}
+ a_{\uparrow -+}^{(\mathrm{S})}\,\,\, \mathcal{M}_{\uparrow
-+}^{(\mathrm{S})}\,\,\, \begin{pmatrix} u_{\uparrow
-+}^{(\mathrm{S})} \\[5pt] v_{\uparrow -+}^{(\mathrm{S})}
\end{pmatrix} \,\,\, \ee^{i\, k_{\uparrow -+}^{(\mathrm{S})}\,
x} \nonumber \\ &\hspace{0.3cm} +  a_{\downarrow
+-}^{(\mathrm{S})} \,\,\, \mathcal{M}_{\downarrow
+-}^{(\mathrm{S})}\,\,\, \begin{pmatrix} u_{\downarrow
+-}^{(\mathrm{S})} \\[5pt] v_{\downarrow +-}^{(\mathrm{S})}
\end{pmatrix}\,\,\, \ee^{i\, k_{\downarrow +-}^{(\mathrm{S})}\,
x} + a_{\downarrow --}^{(\mathrm{S})}\,\,\,
\mathcal{M}_{\downarrow --}^{(\mathrm{S})}\,\,\, \begin{pmatrix}
u_{\downarrow --}^{(\mathrm{S})} \\[5pt] v_{\downarrow
--}^{(\mathrm{S})} \end{pmatrix} \,\,\, \ee^{i\,
k_{\downarrow --}^{(\mathrm{S})}\, x} \,\,\, .
\end{align}
Imposing the matching condition (\ref{eq:SSpConti}) on the
\textit{Ans\"atze}\/ for evanescent-quasiparticle excitations
in the TSF and NSF regions yields a homogeneous system of 8
linear equations for the 8 superposition coefficients
$a_{\sigma\tau\alpha}^{(R)}$. Similarly to the approach
discussed at the end of Sec.~\ref{sec:ISform} for treating the
TSF edge, the unwieldy system of equations arising from
matching at the SS$^\prime$ interface can be recast in an
equivalent $4\times 4$ form;
\begin{equation}\label{eq:sUpEquSSp}
\begin{pmatrix} a_{\downarrow +-}^{(\mathrm{S})} \\[0.1cm]
a_{\downarrow --}^{(\mathrm{S})} \\[0.1cm] a_{\downarrow
++}^{(\mathrm{S'})} \\[0.1cm] a_{\downarrow --}^{(\mathrm{S'})}
\end{pmatrix} = -\left[
\mathcal{D}_\downarrow^{(\mathrm{SS'})} \right]^{-1}
\mathcal{C}_{\downarrow\uparrow}^{(\mathrm{SS'})} \,\,
\begin{pmatrix} a_{\uparrow +-}^{(\mathrm{S})} \\[0.1cm]
a_{\uparrow -+}^{(\mathrm{S})} \\[0.1cm] a_{\uparrow
++}^{(\mathrm{S'})} \\[0.1cm] a_{\uparrow --}^{(\mathrm{S'})}
\end{pmatrix} \,\, , \,\, \left\{
\mathcal{D}_\uparrow^{(\mathrm{SS'})} -
\mathcal{C}_{\uparrow\downarrow}^{(\mathrm{SS'})} \left[
\mathcal{D}_\downarrow^{(\mathrm{SS'})} \right]^{-1}
\mathcal{C}_{\downarrow\uparrow}^{(\mathrm{SS'})} \right\}
\begin{pmatrix} a_{\uparrow +-}^{(\mathrm{S})} \\[0.1cm]
a_{\uparrow -+}^{(\mathrm{S})} \\[0.1cm] a_{\uparrow
++}^{(\mathrm{S'})} \\[0.1cm] a_{\uparrow --}^{(\mathrm{S'})}
\end{pmatrix}  = \begin{pmatrix} 0 \\[0.2cm] 0 \\[0.2cm] 0
\\[0.2cm] 0 \end{pmatrix}\,\, ,
\end{equation}
with the matrices
\begin{subequations}
\begin{align}\label{eq:DupSSp}
\mathcal{D}_\uparrow^{(\mathrm{SS'})} &= \begin{pmatrix}
u_{\uparrow +-}^{(\mathrm{S})} & u_{\uparrow -+}^{(\mathrm{S})}
& -u_{\uparrow ++}^{(\mathrm{S'})} &
-u_{\uparrow --}^{(\mathrm{S'})} \\[0.2cm]
v_{\uparrow +-}^{(\mathrm{S})} & v_{\uparrow -+}^{(\mathrm{S})}
& -v_{\uparrow ++}^{(\mathrm{S'})} &
-v_{\uparrow --}^{(\mathrm{S'})} \\[0.2cm]
k_{\uparrow +-}^{(\mathrm{S})}\, u_{\uparrow +-}^{(\mathrm{S})}
& k_{\uparrow -+}^{(\mathrm{S})}\, u_{\uparrow
-+}^{(\mathrm{S})} & -k_{\uparrow ++}^{(\mathrm{S'})}\,
u_{\uparrow ++}^{(\mathrm{S'})} & -k_{\uparrow
--}^{(\mathrm{S'})}\, u_{\uparrow --}^{(\mathrm{S'})} \\[0.2cm]
k_{\uparrow +-}^{(\mathrm{S})}\, v_{\uparrow +-}^{(\mathrm{S})}
& k_{\uparrow -+}^{(\mathrm{S})}\, v_{\uparrow
-+}^{(\mathrm{S})} & -k_{\uparrow ++}^{(\mathrm{S'})}\,
v_{\uparrow ++}^{(\mathrm{S'})} & -k_{\uparrow
--}^{(\mathrm{S'})}\, v_{\uparrow --}^{(\mathrm{S'})}
\end{pmatrix} \quad , \\[0.2cm] \label{eq:DdownSSp}
\mathcal{D}_\downarrow^{(\mathrm{SS'})} &= \begin{pmatrix}
u_{\downarrow +-}^{(\mathrm{S})} & u_{\downarrow
--}^{(\mathrm{S})} & -u_{\downarrow ++}^{(\mathrm{S'})} &
-u_{\downarrow --}^{(\mathrm{S'})} \\[0.2cm]
v_{\downarrow +-}^{(\mathrm{S})} & v_{\downarrow
--}^{(\mathrm{S})} & -v_{\downarrow ++}^{(\mathrm{S'})} &
-v_{\downarrow --}^{(\mathrm{S'})} \\[0.2cm]
k_{\downarrow +-}^{(\mathrm{S})}\, u_{\downarrow
+-}^{(\mathrm{S})} & k_{\downarrow --}^{(\mathrm{S})}\,
u_{\downarrow --}^{(\mathrm{S})} & -k_{\downarrow
++}^{(\mathrm{S'})}\, u_{\downarrow ++}^{(\mathrm{S'})} &
-k_{\downarrow --}^{(\mathrm{S'})}\, u_{\downarrow
--}^{(\mathrm{S'})} \\[0.2cm] k_{\downarrow +-}^{(\mathrm{S})}\,
v_{\downarrow +-}^{(\mathrm{S})} & k_{\downarrow
--}^{(\mathrm{S})}\, v_{\downarrow --}^{(\mathrm{S})} &
-k_{\downarrow ++}^{(\mathrm{S'})}\, v_{\downarrow
++}^{(\mathrm{S'})} & -k_{\downarrow --}^{(\mathrm{S'})}\,
v_{\downarrow --}^{(\mathrm{S'})}\end{pmatrix} \quad , \\[0.2cm]
\mathcal{C}_{\downarrow\uparrow}^{(\mathrm{SS'})} &=
\begin{pmatrix} \mathcal{J}_{\uparrow +-}^{(\mathrm{S})}
\begin{pmatrix} u_{\uparrow +-}^{(\mathrm{S})} \\[0.1cm]
v_{\uparrow +-}^{(\mathrm{S})}\end{pmatrix} &
\mathcal{J}_{\uparrow -+}^{(\mathrm{S})} \begin{pmatrix}
u_{\uparrow -+}^{(\mathrm{S})} \\[0.1cm] v_{\uparrow
-+}^{(\mathrm{S})} \end{pmatrix} & - \mathcal{J}_{\uparrow
++}^{(\mathrm{S'})} \begin{pmatrix} u_{\uparrow
++}^{(\mathrm{S'})} \\[0.1cm] v_{\uparrow ++}^{(\mathrm{S'})}
\end{pmatrix} & - \mathcal{J}_{\uparrow --}^{(\mathrm{S'})}
\begin{pmatrix} u_{\uparrow --}^{(\mathrm{S'})} \\[0.1cm]
v_{\uparrow --}^{(\mathrm{S'})} \end{pmatrix} \\[0.2cm]
k_{\uparrow +-}^{(\mathrm{S})}\, \mathcal{J}_{\uparrow
+-}^{(\mathrm{S})}\begin{pmatrix} u_{\uparrow +-}^{(\mathrm{S})}
\\[0.1cm] v_{\uparrow +-}^{(\mathrm{S})}\end{pmatrix} &
k_{\uparrow -+}^{(\mathrm{S})}\, \mathcal{J}_{\uparrow
-+}^{(\mathrm{S})}\begin{pmatrix} u_{\uparrow -+}^{(\mathrm{S})}
\\[0.1cm] v_{\uparrow-+}^{(\mathrm{S})} \end{pmatrix} & -
k_{\uparrow ++}^{(\mathrm{S'})}\, \mathcal{J}_{\uparrow
++}^{(\mathrm{S'})} \begin{pmatrix} u_{\uparrow
++}^{(\mathrm{S'})} \\[0.1cm] v_{\uparrow ++}^{(\mathrm{S'})}
\end{pmatrix} & - k_{\uparrow --}^{(\mathrm{S'})}\,
\mathcal{J}_{\uparrow --}^{(\mathrm{S'})} \begin{pmatrix}
u_{\uparrow --}^{(\mathrm{S'})} \\[0.1cm] v_{\uparrow
--}^{(\mathrm{S'})} \end{pmatrix}\end{pmatrix} , \\[0.2cm]
\mathcal{C}_{\uparrow\downarrow}^{(\mathrm{SS'})} &=
\begin{pmatrix} \mathcal{J}_{\downarrow +-}^{(\mathrm{S})}
\begin{pmatrix} u_{\downarrow +-}^{(\mathrm{S})} \\[0.1cm]
v_{\downarrow +-}^{(\mathrm{S})}\end{pmatrix} &
\mathcal{J}_{\downarrow --}^{(\mathrm{S})} \begin{pmatrix}
u_{\downarrow --}^{(\mathrm{S})} \\[0.1cm] v_{\downarrow
--}^{(\mathrm{S})} \end{pmatrix} & - \mathcal{J}_{\downarrow
++}^{(\mathrm{S'})} \begin{pmatrix} u_{\downarrow
++}^{(\mathrm{S'})} \\[0.1cm] v_{\downarrow ++}^{(\mathrm{S'})}
\end{pmatrix} & - \mathcal{J}_{\downarrow --}^{(\mathrm{S'})}
\begin{pmatrix} u_{\downarrow --}^{(\mathrm{S'})} \\[0.1cm]
v_{\downarrow --}^{(\mathrm{S'})} \end{pmatrix} \\[0.2cm]
k_{\downarrow +-}^{(\mathrm{S})}\, \mathcal{J}_{\downarrow
+-}^{(\mathrm{S})}\begin{pmatrix} u_{\downarrow
+-}^{(\mathrm{S})} \\[0.1cm] v_{\downarrow +-}^{(\mathrm{S})}
\end{pmatrix} & k_{\downarrow --}^{(\mathrm{S})}\,
\mathcal{J}_{\downarrow --}^{(\mathrm{S})}\begin{pmatrix}
u_{\downarrow --}^{(\mathrm{S})} \\[0.1cm]
v_{\downarrow--}^{(\mathrm{S})} \end{pmatrix} & -
k_{\downarrow ++}^{(\mathrm{S'})}\, \mathcal{J}_{\downarrow
++}^{(\mathrm{S'})} \begin{pmatrix} u_{\downarrow
++}^{(\mathrm{S'})} \\[0.1cm] v_{\downarrow ++}^{(\mathrm{S'})}
\end{pmatrix} & - k_{\downarrow --}^{(\mathrm{S'})}\,
\mathcal{J}_{\downarrow --}^{(\mathrm{S'})} \begin{pmatrix}
u_{\downarrow --}^{(\mathrm{S'})} \\[0.1cm] v_{\downarrow
--}^{(\mathrm{S'})} \end{pmatrix}\end{pmatrix} .
\end{align}
\end{subequations}
Alternatively to (\ref{eq:sUpEquSSp}), the original $8\times 8$
system of linear equations can be formally expressed in terms of
an equivalent other set of $4\times 4$ relations 
\begin{equation}\label{eq:sDownEquSSp}
\begin{pmatrix} a_{\uparrow +-}^{(\mathrm{S})} \\[0.1cm]
a_{\uparrow -+}^{(\mathrm{S})} \\[0.1cm] a_{\uparrow
++}^{(\mathrm{S'})} \\[0.1cm] a_{\uparrow --}^{(\mathrm{S'})}
\end{pmatrix} = -\left[ \mathcal{D}_\uparrow^{(\mathrm{SS'})}
\right]^{-1} \mathcal{C}_{\uparrow\downarrow}^{(\mathrm{SS'})}
\,\, \begin{pmatrix} a_{\downarrow +-}^{(\mathrm{S})} \\[0.1cm]
a_{\downarrow --}^{(\mathrm{S})} \\[0.1cm] a_{\downarrow
++}^{(\mathrm{S'})} \\[0.1cm] a_{\downarrow --}^{(\mathrm{S'})}
\end{pmatrix} \,\, , \,\, \left\{
\mathcal{D}_\downarrow^{(\mathrm{SS'})} -
\mathcal{C}_{\downarrow\uparrow}^{(\mathrm{SS'})} \left[
\mathcal{D}_\uparrow^{(\mathrm{SS'})} \right]^{-1}
\mathcal{C}_{\uparrow\downarrow}^{(\mathrm{SS'})} \right\}
\begin{pmatrix} a_{\downarrow +-}^{(\mathrm{S})} \\[0.1cm]
a_{\downarrow --}^{(\mathrm{S})} \\[0.1cm] a_{\downarrow
++}^{(\mathrm{S'})} \\[0.1cm] a_{\downarrow --}^{(\mathrm{S'})}
\end{pmatrix}  = \begin{pmatrix} 0 \\[0.2cm] 0 \\[0.2cm] 0
\\[0.2cm] 0 \end{pmatrix}\,\, .
\end{equation}

The form of the matrices $\mathcal{D}_\sigma^{(\mathrm{SS'})}$
could be interpreted as arising from matching conditions for
large Nambu-spinor components of the isolated spin-$\sigma$
subsystems, with each of these constituting a junction between
chiral-\textit{p}-wave superfluids with the same chirality on
both sides of the interface but different pair-potential
magnitudes. This analogy is most immediate for the
spin-$\uparrow$ sector whose effective chemical potential $\mu -
\nu_\uparrow^{(R)}$ is positive in both the S and S$^\prime$
regions, and for which the interface therefore constitutes a
genuine Josephson junction. In contrast, the fact that $\mu -
\nu_\downarrow^{(\mathrm{S})}< 0 < \mu -
\nu_\downarrow^{(\mathrm{S}^\prime)}$ suggests that the TSF-NSF
interface acts like a wall for spin-$\downarrow$ quasiparticles
from the NSF region, raising the possibility for a Majorana
edge state to emerge.

The following subsection \ref{sec:SSPdecoupled} formalises the
description of the SS$^\prime$ interface in terms of
chiral-\textit{p}-wave junctions of fully separated spin
subsystems. To provide a reference point for comparison, as well
as further relevant background information, we derive the
Andreev bound states at the most general realization of an
interface between two arbitrary spinless chiral-\textit{p}-wave
systems in Appendix~\ref{app:chiPwave}. Results presented there
generalize those given in related previous works~\cite{Ho1984,
Matsumoto1999,Barash2001,Kwon2004,Samokhin2012} to the situation
where the order-parameter magnitudes are different in the two
regions.

While it may be tempting to limit considerations to the matrices
$\mathcal{D}_\sigma^{(\mathrm{SS'})}$ representing individual
spin-$\sigma$ subsystems, such an approach neglects the coupling
between spin-$\uparrow$ and spin-$\downarrow$ sectors embodied
in the matrices $\mathcal{C}_{\sigma\bar\sigma}^{(\mathrm{SS'}
)}$. In effect, our SS$^\prime$ hybrid system should be most
appropriately thought of as a Josephson junction between two
two-band superconductors where the two bands are distinguished
by the spin projection $\sigma\in \{\uparrow, \downarrow\}$.
Existing theoretical descriptions of two-band-superconductor
junctions rely heavily on numerics~\cite{Sasaki2020}. In
sections \ref{sec:SSPdownC} and \ref{sec:SSPupC}, we develop
approximate analytical methods to explore whether and how the
residual coupling between opposite-spin sectors modifies the
predictions obtained within the simple picture from
Sec.~\ref{sec:SSPdecoupled}. Physical consequences for the
Andreev-bound-state spectrum at the TSF-NSF interface are
discussed in detail in Sec.~\ref{sec:resInter}.

\subsubsection{Description in terms of completely decoupled spin
subsystems}\label{sec:SSPdecoupled}

On the most elementary level, the energies of
interface-localized Andreev bound states within individual
spin-$\sigma$ sectors can be found as solutions of the
characteristic equations
\begin{subequations}
\begin{align}\label{eq:SSpSepSup}
\mathrm{det}\left(\mathcal{D}_\uparrow^{(\mathrm{SS'})}\right)
&\approx 4\, \big( k_{\mathrm{F}\uparrow}^{(\mathrm{SS'})}
\big)^2 \left( u_{\uparrow+-}^{(\mathrm{S})}\,
v_{\uparrow--}^{(\mathrm{S'})} - u_{\uparrow
--}^{(\mathrm{S'})}\, v_{\uparrow+-}^{(\mathrm{S})} \right)
\left( u_{\uparrow-+}^{(\mathrm{S})}\, v_{\uparrow
++}^{(\mathrm{S'})} - u_{\uparrow++}^{(\mathrm{S'})}\,
v_{\uparrow-+}^{(\mathrm{S})} \right) = 0 \quad , \\
\mathrm{det}\left(\mathcal{D}_\downarrow^{(\mathrm{SS'})}\right)
&\approx \left[ \big| k_{\mathrm{F}\downarrow}^{(\mathrm{S})}
\big|^2 + \big( k_{\mathrm{F}\downarrow}^{(\mathrm{S'})} \big)^2
\right] \left( u_{\downarrow+-}^{(\mathrm{S})}\,
v_{\downarrow--}^{(\mathrm{S})} - u_{\downarrow
--}^{(\mathrm{S})}\, v_{\downarrow+-}^{(\mathrm{S})} \right)
\left( u_{\downarrow--}^{(\mathrm{S'})}\,
v_{\downarrow++}^{(\mathrm{S'})} - u_{\downarrow
++}^{(\mathrm{S'})}\, v_{\downarrow--}^{(\mathrm{S'})} \right)
= 0 \quad . \label{eq:SSpSepSdown}
\end{align}
\end{subequations}
Here the determinants of matrices
$\mathcal{D}_\uparrow^{(\mathrm{SS'})}$ and
$\mathcal{D}_\downarrow^{(\mathrm{SS'})}$ given explicitly in
Eqs.~(\ref{eq:DupSSp}) and (\ref{eq:DdownSSp}), respectively,
have been calculated to leading order in the Andreev
approximation (\ref{eq:kAA}) where
\begin{subequations}
\begin{align}\label{eq:kFSSdef}
k_{\uparrow\tau\alpha}^{(R)} &\approx \alpha\, k_{\mathrm{F}
\uparrow}^{(\mathrm{SS'})} \quad \mbox{with}\quad k_{\mathrm{F}
\uparrow}^{(\mathrm{SS'})}\equiv\sqrt{2 m (\mu + h)/\hbar^2}
\quad , \\[2pt] \label{eq:kFdownAA}
k_{\downarrow\tau\alpha}^{(\mathrm{S'})} &\approx \alpha\,
k_{\mathrm{F}\downarrow}^{(\mathrm{S'})} \quad\mbox{and}\quad
k_{\downarrow\tau\alpha}^{(\mathrm{S})} \approx \alpha\, i\,
\big|k_{\mathrm{F}\downarrow}^{(\mathrm{S})}\big| \quad .
\end{align}
\end{subequations}

Using also the Andreev-approximation expressions (\ref{eq:uvAA})
for Nambu-spinor components in the
separated-spin-$\uparrow$-sector characteristic equation
(\ref{eq:SSpSepSup}) yields the $Z=0$, $\gamma^{(\mathrm{S})}=
\gamma^{(\mathrm{S'})}$ limit of the general
spinless-chiral-\textit{p}-wave-junction characteristic equation
derived in Appendix~\ref{app:chiPwave} [see
Eq.~(\ref{eq:simpPwaveSec})], which reads explicitly
\begin{equation}\label{eq:sUpSSpABS}
\cos\left( \theta_\uparrow^{(\mathrm{S'})} +
\theta_\uparrow^{(\mathrm{S})}\right) = \cos\left(
\varphi^{(\mathrm{S'})} - \varphi^{(\mathrm{S})}\right) \quad .
\end{equation}
Thus, within the Andreev-approximation treatment, the completely
decoupled spin-$\uparrow$ sector of the considered realization
of a TSF-NSF interface constitutes a fully transparent Josephson
junction between equal-chirality \textit{p}-wave superfluids
with different pair-potential magnitude.
Appendix~\ref{app:chiPwave} provides a more detailed
discussion of such a system's physical properties, including a
comparison with those of a conventional \text{s}-wave junction.

The fact that $\mu-\nu_\downarrow^{(\mathrm{S})} < 0$ implies
that the separated-spin-$\downarrow$-sector characteristic
equation (\ref{eq:SSpSepSdown}) can only be satisfied when the
relation
\begin{equation}\label{eq:MajAppear}
u_{\downarrow--}^{(\mathrm{S'})}\, v_{\downarrow
++}^{(\mathrm{S'})} - u_{\downarrow++}^{(\mathrm{S'})}\,
v_{\downarrow--}^{(\mathrm{S'})} = 0
\end{equation}
holds, which is the defining equation for the surface bound
state of an unconventional superconductor~\cite{Kashiwaya1995,
Kashiwaya2000}, including the Majorana edge mode in a
chiral-\textit{p}-wave superfluid~\cite{Honerkamp1998}. In
particular, inserting the Andreev-approximation expressions
(\ref{eq:uvAA}) for Nambu-spinor amplitudes appearing in
Eq.~(\ref{eq:MajAppear}) transforms the latter into
\begin{align}\label{eq:TSFsecSSpAA}
\cos\left[\theta_\downarrow^{(\mathrm{S'})} + \sgn(E)\,
\vartheta_{k_y\,\downarrow}^{(\mathrm{S'})} \,\right] = 0
\quad .
\end{align}
We discuss in more detail in Appendix~\ref{app:chiPwave} how
the Majorana-mode energy dispersion emerges as the solution of
characteristic equations that are of the form
(\ref{eq:TSFsecSSpAA}), and the Majorana-fermion character of
the corresponding eigenstates is demonstrated in
Appendix~\ref{app:Majorana}. Thus the intuitive expectation that
a TSF-NSF interface constitutes a boundary for the
spin-$\downarrow$-sector quasiparticles in the NSF region is
borne out within the separated-spin-sector description.

\subsubsection{Opposite-spin-coupling modifications for the
spin-$\downarrow$ subsystem}\label{sec:SSPdownC}

We base our investigation of modifications generated in the
spin-$\downarrow$-sector characteristic equation by the
coupling to the spin-$\uparrow$ degrees of freedom on the
relations (\ref{eq:sDownEquSSp}). Requiring that
the system of linear equations on the right-hand side of
Eq.~(\ref{eq:sDownEquSSp}) has a nontrivial solution leads to
the characteristic equation
\begin{equation}\label{eq:TSFsec4x4}
\mathrm{det} \left( \mathcal{D}_\downarrow^{(\mathrm{SS'})} -
\mathcal{C}_{\downarrow\uparrow}^{(\mathrm{SS'})} \left[
\mathcal{D}_\uparrow^{(\mathrm{SS'})} \right]^{-1}
\mathcal{C}_{\uparrow\downarrow}^{(\mathrm{SS'})}\right) = 0
\quad .
\end{equation}
As the spin-$\uparrow$ sector is deeply in the BCS regime and
$h\sim\mu$ as per Eq.~(\ref{eq:TSFtoNSF}), it is possible to
neglect $|\Delta^{(R)}|/h$ corrections in the matrices
$\mathcal{J}_{\uparrow\tau\alpha}^{(R)}$. Using also the
approximation (\ref{eq:kFSSdef}), we find
\begin{equation}\label{eq:spinDownMat}
\mathcal{C}_{\downarrow\uparrow}^{(\mathrm{SS'})} \left[
\mathcal{D}_\uparrow^{(\mathrm{SS'})} \right]^{-1} \approx
\frac{\lambda}{2 h} \begin{pmatrix} - k_y & 0 & i & 0 \\[2pt]
0 & k_y & 0 & i \\[2pt] i \big( k_{\mathrm{F}
\uparrow}^{(\mathrm{SS'})}\big)^2 & 0 & - k_y & 0 \\[2pt] 0 & i
\big( k_{\mathrm{F}\uparrow}^{(\mathrm{SS'})}\big)^2 & 0 & k_y
\end{pmatrix}\quad .
\end{equation}
Utilizing the expression (\ref{eq:spinDownMat}) in the term
providing a correction to
$\mathcal{D}_\downarrow^{(\mathrm{SS'})}$ in
(\ref{eq:TSFsec4x4}) yields
\begin{align}\label{eq:GenCorrSSp}
\mathcal{C}_{\downarrow\uparrow}^{(\mathrm{SS'})} \left[
\mathcal{D}_\uparrow^{(\mathrm{SS'})} \right]^{-1}
\mathcal{C}_{\uparrow\downarrow}^{(\mathrm{SS'})} =
-\sqrt{\frac{m \lambda^2}{\hbar^2 h}}\,\,
\mathcal{A}_\downarrow^{(\mathrm{SS'})} - \frac{m
\lambda^2}{\hbar^2 h}\,\,\mathcal{B}_\downarrow^{(\mathrm{SS'})
} \quad .
\end{align}
The relation $h \sim \mu$ noted already above also implies that
the spin-$\downarrow$ wave numbers $k_{\downarrow\tau
\alpha}^{(R)}$ appearing in the $4\times4$ matrices
$\mathcal{D}_\downarrow^{(\mathrm{SS'})}$,
$\mathcal{A}_\downarrow^{(\mathrm{SS'})}$ and
$\mathcal{B}_\downarrow^{(\mathrm{SS'})}$ are small quantities.
Specifically, we have $k_{\downarrow\tau\alpha}^{(R)} =
\mathcal{O} \bigg( \sqrt{m_\downarrow^{(R)} \big| \Delta^{(R)}
\big|^2/(\hbar^2 h)} \,\bigg)$ and, consequently, $k_{\downarrow
\tau\alpha}^{(R)}/k_{\mathrm{F}\uparrow}^{(\mathrm{SS'})} =
\mathcal{O}\big(\big|\Delta^{(R)}\big|/h\big)$. Neglecting terms
$\mathcal{O}\big(\big|\Delta^{(R)}\big|/h\big)$ as well as
$\mathcal{O}\big(\big[k_y/k_{\downarrow\tau\alpha}^{(R)}\big]^2
\big)$, consistent with the Andreev approximation as applied to
the spin-$\uparrow$ sector [see the discussion preceding
Eq.~(\ref{eq:spinDownMat})], then yields
\begin{subequations}
\begin{align}\label{eq:Adown}
&\mathcal{A}_\downarrow^{(\mathrm{SS'})} \approx i\,\,
\sqrt{\frac{m}{\hbar^2 h}}\,\,\,\begin{pmatrix} 0 & 0 & 0 & 0
\\[0.2cm] 0 & 0 & 0 & 0 \\[0.2cm] \Delta^{(\mathrm{S})} \,
v_{\downarrow+-}^{(\mathrm{S})} & \Delta^{(\mathrm{S})} \,
v_{\downarrow--}^{(\mathrm{S})} & -\Delta^{(\mathrm{S'})} \,
v_{\downarrow++}^{(\mathrm{S'})} & -\Delta^{(\mathrm{S'})} \,
v_{\downarrow--}^{(\mathrm{S'})} \\[0.2cm]
\big(\Delta^{(\mathrm{S})}\big)^* \,
u_{\downarrow+-}^{(\mathrm{S})} & \big( \Delta^{(\mathrm{S})}
\big)^* \, u_{\downarrow--}^{(\mathrm{S})} &
-\big(\Delta^{(\mathrm{S'})}\big)^* \,
u_{\downarrow++}^{(\mathrm{S'})} & -\big(\Delta^{(\mathrm{S'})}
\big)^*\,u_{\downarrow--}^{(\mathrm{S'})} \end{pmatrix}\,\, ,
\\[0.2cm] \label{eq:Bdown}
&\mathcal{B}_\downarrow^{(\mathrm{SS'})}\approx\nonumber\\[5pt]
&\begin{pmatrix} 0 & 0 & 0 & 0 \\[0.4cm] 0 & 0 & 0 & 0 \\[0.4cm]
\begin{matrix} u_{\downarrow+-}^{(\mathrm{S})}\,\Biggl\{
k_{\downarrow+-}^{(\mathrm{S})} - i\, k_y \\ \hfill \times
\Biggl[1 - \biggl(\frac{k_{\downarrow+-}^{(\mathrm{S})}}
{k_{\mathrm{F}\uparrow}^{(\mathrm{SS'})}}\biggr)^2\Biggr]
\Biggr\}\end{matrix} & \begin{matrix}
u_{\downarrow--}^{(\mathrm{S})}\,\Biggl\{
k_{\downarrow--}^{(\mathrm{S})} - i\, k_y \\ \hfill \times
\Biggl[1 - \biggl(\frac{k_{\downarrow--}^{(\mathrm{S})}}
{k_{\mathrm{F}\uparrow}^{(\mathrm{SS'})}}\biggr)^2\Biggl]
\Bigg\}\end{matrix} & \begin{matrix}
-u_{\downarrow++}^{(\mathrm{S'})}\,\Biggl\{
k_{\downarrow++}^{(\mathrm{S'})} - i\, k_y \\ \hfill \times
\Biggl[1 - \biggl(\frac{k_{\downarrow++}^{(\mathrm{S'})}}
{k_{\mathrm{F}\uparrow}^{(\mathrm{SS'})}}\biggr)^2\Biggr]
\Biggr\}\end{matrix} & \begin{matrix}
-u_{\downarrow--}^{(\mathrm{S'})}\,\Biggl\{
k_{\downarrow--}^{(\mathrm{S'})} - i\, k_y \\ \hfill \times
\Biggl[1 - \biggl(\frac{k_{\downarrow--}^{(\mathrm{S'})}}
{k_{\mathrm{F}\uparrow}^{(\mathrm{SS'})}}\biggr)^2\Biggr]
\Biggr\}\end{matrix} \\[1.2cm] \begin{matrix}
v_{\downarrow+-}^{(\mathrm{S})}\,\Biggl\{
k_{\downarrow+-}^{(\mathrm{S})} + i\, k_y \\ \hfill \times
\Biggl[1 - \biggl(\frac{k_{\downarrow+-}^{(\mathrm{S})}}
{k_{\mathrm{F}\uparrow}^{(\mathrm{SS'})}}\biggr)^2\Biggr]
\Biggr\}\end{matrix} & \begin{matrix}
v_{\downarrow--}^{(\mathrm{S})}\,\Biggl\{
k_{\downarrow--}^{(\mathrm{S})} +\, i\, k_y \\ \hfill \times
\Biggl[1 - \biggl(\frac{k_{\downarrow--}^{(\mathrm{S})}}
{k_{\mathrm{F}\uparrow}^{(\mathrm{SS'})}}\biggr)^2\Biggr]
\Biggr\}\end{matrix} & \begin{matrix}
-v_{\downarrow++}^{(\mathrm{S'})}\,\Biggl\{
k_{\downarrow++}^{(\mathrm{S'})} +\, i\, k_y \\ \hfill \times
\Biggl[1 - \biggl(\frac{k_{\downarrow++}^{(\mathrm{S'})}}
{k_{\mathrm{F}\uparrow}^{(\mathrm{SS'})}}\biggr)^2\Biggr]
\Biggr\}\end{matrix} & \begin{matrix}
-v_{\downarrow--}^{(\mathrm{S'})}\,\Biggl\{
k_{\downarrow--}^{(\mathrm{S'})} +\, i\, k_y \\ \hfill \times
\Biggl[1 - \biggl(\frac{k_{\downarrow--}^{(\mathrm{S'})}}
{k_{\mathrm{F}\uparrow}^{(\mathrm{SS'})}}\biggr)^2\Biggr]
\Biggr\}\end{matrix} \end{pmatrix} \,\, . 
\end{align}
\end{subequations}

Subleading-order corrections to factors multiplying $k_y$ have
been retained in $\mathcal{B}_\downarrow^{(\mathrm{SS'})}$ for
illustration, as these contribute a leading-order correction in
the limit of small spin-orbit coupling. The nonzero entries of
$\mathcal{A}_\downarrow^{(\mathrm{SS'})}$ as given in
Eq.~(\ref{eq:Adown}) have a magnitude $\mathcal{O}\bigg(\sqrt{m
\big|\Delta^{(R)}\big|^2/(\hbar^2 h)}\,\bigg)$, which is the
same as that of the $k_{\downarrow\tau\alpha}^{(R)}$ prefactors
of entries in the bottom two rows of
$\mathcal{D}_\downarrow^{(\mathrm{SS'})}$ [see
Eq.~(\ref{eq:DdownSSp})]. However, the additional factor
$\sqrt{m\lambda^2/(\hbar^2 h)}\ll 1$ in front of
$\mathcal{A}_\downarrow^{(\mathrm{SS'})}$ in
Eq.~(\ref{eq:GenCorrSSp}) renders this term's contribution
negligible within the Andreev approximation [see
Eq.~(\ref{eq:AAexplicit}) and discussion below]. In contrast,
the correction term involving the matrix
$\mathcal{B}_\downarrow^{(\mathrm{SS'})}$ in
Eq.~(\ref{eq:GenCorrSSp}) formally constitutes a leading-order
correction in small spin-orbit-coupling magnitude to
$\mathcal{D}_\downarrow^{(\mathrm{SS'})}$. Retaining this term
when evaluating (\ref{eq:TSFsec4x4}) would yield the
characteristic equation
\begin{align}\label{eq:SSpDownL2}
\cos\left[\theta_\downarrow^{(\mathrm{S'})} + \sgn(E)\, \left(
\vartheta_{k_y\,\downarrow}^{(\mathrm{S'})} + \frac{\lambda^2}{2
h^2}\,k_{\mathrm{F}\downarrow}^{(\mathrm{S'})}\, k_y\right)
\right] = 0 \quad ,
\end{align}
which is of the same general form as Eq.~(\ref{eq:TSFsecAA})
found for the I-TSF edge state. However, because of the
parametric smallness of
$k_{\mathrm{F}\downarrow}^{(\mathrm{S'})} = \mathcal{O} \bigg(
\sqrt{m_\downarrow^{(\mathrm{S'})} \big| \Delta^{(\mathrm{S'})}
\big|^2/(\hbar^2 h)} \bigg)$ within our particular realization
of the TSF-NSF interface, overall consistency requires that
the $\lambda^2$-dependent correction in Eq.~(\ref{eq:SSpDownL2})
is also neglected. As a consequence, to leading order in the
Andreev approximation, the coupling to the spin-$\uparrow$
subsystem does not affect the properties of the
spin-$\downarrow$ Andreev bound state formed at the SS$^\prime$
interface, and the characteristic equation yielding its energy
dispersion is given by the result Eq.~(\ref{eq:TSFsecSSpAA})
obtained in the limit where the spin subsectors are considered
to be uncoupled.

\subsubsection{Opposite-spin-coupling modifications for the
spin-$\uparrow$ subsystem}\label{sec:SSPupC}

We now turn to discussing properties of the Josephson junction
realized in the spin-$\uparrow$ sector, utilizing the relations
shown in Eqs.~(\ref{eq:sUpEquSSp}). The characteristic equation
for interface-localized bound states is
\begin{equation}\label{eq:UpSec4x4}
\mathrm{det} \left( \mathcal{D}_\uparrow^{(\mathrm{SS'})} -
\mathcal{C}_{\uparrow\downarrow}^{(\mathrm{SS'})} \left[
\mathcal{D}_\downarrow^{(\mathrm{SS'})} \right]^{-1}
\mathcal{C}_{\downarrow\uparrow}^{(\mathrm{SS'})}\right) = 0
\quad .
\end{equation}
As we are interested in obtaining corrections to the
Andreev-approximated form of the characteristic equation
(\ref{eq:SSpSepSup}), we can neglect $|\Delta^{(R)}|/h$
corrections in the matrices $\mathcal{J}_{\downarrow\tau
\alpha}^{(R)}$ entering $\mathcal{C}_{\uparrow
\downarrow}^{(\mathrm{SS'})}$. A straightforward calculation
using the relations from Eq.~(\ref{eq:kFdownAA}) then yields
\begin{align}\label{eq:spinUpMat}
\mathcal{C}_{\uparrow\downarrow}^{(\mathrm{SS'})} \left[
\mathcal{D}_\downarrow^{(\mathrm{SS'})} \right]^{-1} &\approx
\frac{\lambda}{2 h} \begin{pmatrix} k_y & 0 & i & 0 \\[2pt] 0 &
-k_y & 0 & i \\[2pt] 0 & 0 & \big|k_{\mathrm{F}
\downarrow}^{(\mathrm{S})}\big| + k_y & 0 \\[2pt] 0 & 0 & 0 &
\big|k_{\mathrm{F}\downarrow}^{(\mathrm{S})}\big| - k_y
\end{pmatrix} \nonumber \\[5pt] & \hspace{2.5cm} +\,\,
\frac{\lambda\, k_{\mathrm{F}\downarrow}^{(\mathrm{S'})}}{2 h}
\begin{pmatrix} 0 & 0 & 0 & 0 \\[0.2cm] 0 & 0 & 0 & 0 \\[0.2cm]
\big|k_{\mathrm{F}\downarrow}^{(\mathrm{S})}\big|\, a &
\big|k_{\mathrm{F}\downarrow}^{(\mathrm{S})}\big|\, b & -i\, a
& -i\, b \\[0.2cm] \big|k_{\mathrm{F}\downarrow}^{(\mathrm{S})}
\big|\, c & -\big|k_{\mathrm{F}\downarrow}^{(\mathrm{S})}\big|
\, a & -i\, c & i\, a \end{pmatrix}\quad ,
\end{align}
with the abbreviations
\begin{align}\label{eq:abcDef}
a &= \frac{u_{\downarrow--}^{(\mathrm{S'})}\, v_{\downarrow
++}^{(\mathrm{S'})} + u_{\downarrow++}^{(\mathrm{S'})}\,
v_{\downarrow--}^{(\mathrm{S'})}}{u_{\downarrow--}^{(\mathrm{S'}
)}\, v_{\downarrow++}^{(\mathrm{S'})} - u_{\downarrow
++}^{(\mathrm{S'})}\, v_{\downarrow--}^{(\mathrm{S'})}}\,\,\, ,
\,\,\, b = - \frac{2\, u_{\downarrow--}^{(\mathrm{S'})}\,
u_{\downarrow++}^{(\mathrm{S'})}}{u_{\downarrow--}^{(\mathrm{S'}
)}\, v_{\downarrow++}^{(\mathrm{S'})} - u_{\downarrow
++}^{(\mathrm{S'})}\, v_{\downarrow--}^{(\mathrm{S'})}}\,\,\, ,
\,\,\, c = \frac{2\, v_{\downarrow++}^{(\mathrm{S'})}\,
v_{\downarrow--}^{(\mathrm{S'})}}{u_{\downarrow--}^{(\mathrm{S'}
)}\, v_{\downarrow++}^{(\mathrm{S'})} - u_{\downarrow
++}^{(\mathrm{S'})}\, v_{\downarrow--}^{(\mathrm{S'})}}\,\,\, . 
\end{align}

The factors $a$, $b$ and $c$ entering Eq.~(\ref{eq:spinUpMat})
depend on the unknown energy of the spin-$\uparrow$-sector
Andreev bound state via the $E$-dependence of Nambu-spinor
components [see general expressions from Eqs.~(\ref{eq:exactUV})
or their Andreev-approximated forms (\ref{eq:uvAA})]. Also, the
denominator appearing in each of these quantities [see
Eq.~(\ref{eq:abcDef})] coincides with the expression on the
left-hand side of Eq.~(\ref{eq:MajAppear}) that is the
characteristic equation for the energy of the
spin-$\downarrow$-sector Majorana edge state. As a result, a
divergence occurs when $E$ coincides with the spin-$\downarrow$
Majorana-edge-state dispersion. We therefore limit further
discussion to the range of $k_y$ values within which the energy
of the spin-$\uparrow$-sector Andreev bound state is
well-separated from that of the emergent Majorana mode in the
spin-$\downarrow$ sector, as there the quantities $a$, $b$ and
$c$ remain well-defined and all have a magnitude of
$\mathcal{O}(1)$. Due to the parametric smallness of
spin-$\downarrow$ wave vectors, the contribution of the second
term in Eq.~(\ref{eq:spinUpMat}) then becomes negligible.

Using the first term on the right-hand side of
Eq.~(\ref{eq:spinUpMat}) for $\mathcal{C}_{\uparrow
\downarrow}^{(\mathrm{SS'})} \left[
\mathcal{D}_\downarrow^{(\mathrm{SS'})} \right]^{-1}$ in the
characteristic equation (\ref{eq:UpSec4x4}) turns out to only
yield corrections that are small within the Andreev
approximation for spin-$\uparrow$-sector quantities. Hence, at
least to that level of approximation and for the assumed range
of $k_y$ where $E$ is far from the Majorana-edge-state energy of
the spin-$\downarrow$ sector, the description of the
spin-$\uparrow$ system within the completely decoupled limit
remains unmodified.

\section{Results \& discussion I: Edge states at I-NSF
and I-TSF interfaces}
\label{sec:resSurf}

We have derived characteristic equations for the
Andreev-bound-state energies at the edge of a spin-orbit-coupled
polarized 2D Fermi superfluid for when it is in the NSF phase
[Eq.~(\ref{eq:NSFsecAA})] or the TSF phase
[Eq.~(\ref{eq:TSFsecAA})]. These results are accurate to leading
order in the small parameter $m\lambda^2/(\hbar^2 h)$, as our
formalism fundamentally relies on the assumption of small-enough
spin-orbit-coupling magnitude. We also employed the familiar
Andreev approximation~\cite{Andreev1964}, which for our system
of interest implies the relation (\ref{eq:AAexplicit}). We now
present solutions of these characteristic equations and discuss
physical properties of the associated Andreev bound states.

\begin{figure}[t]
\centerline{%
\includegraphics[width=0.9\textwidth]{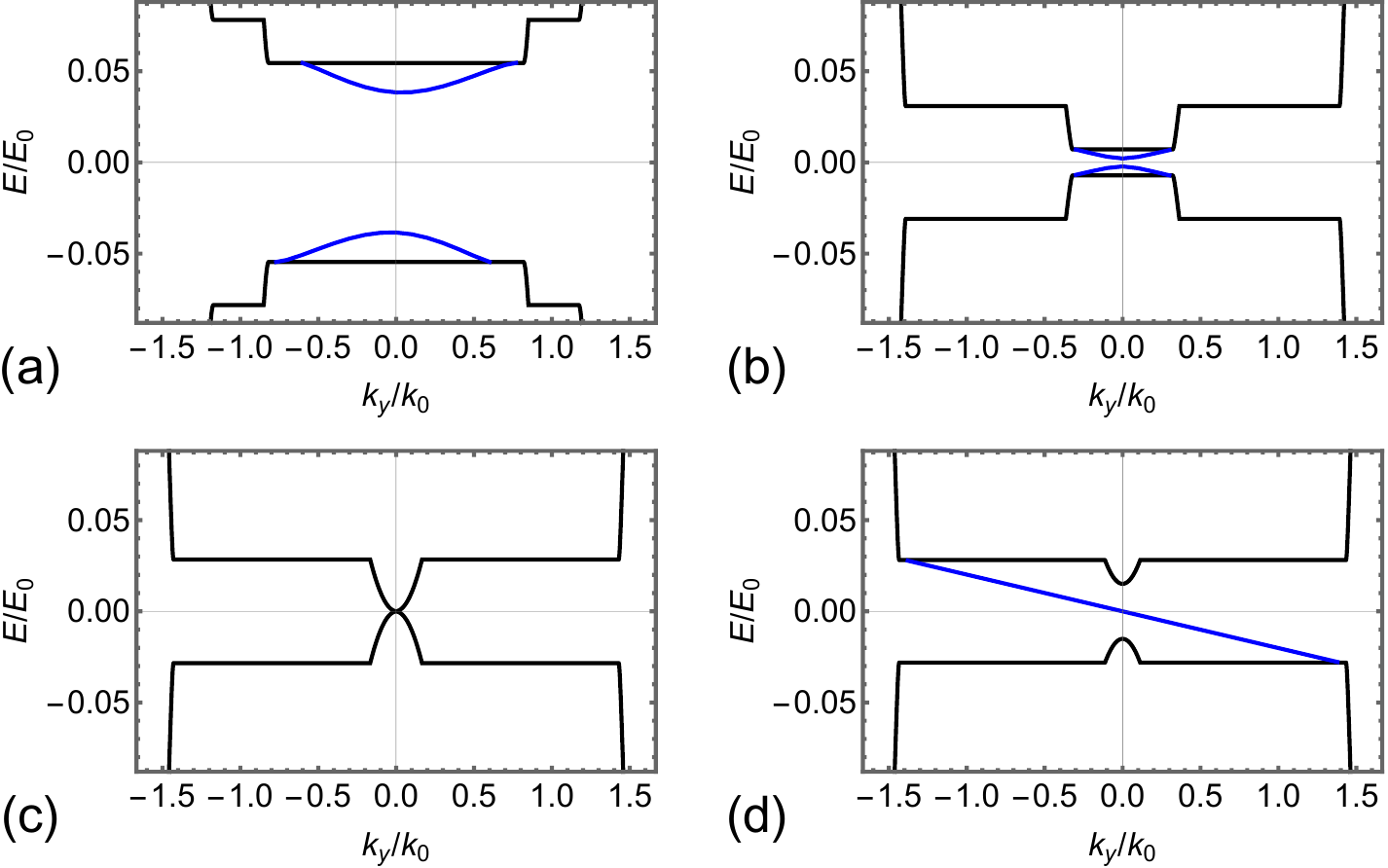}
}%
\caption{\label{fig:ISenergy}%
Energy dispersion of Andreev bound states at an IS interface
oriented parallel to the $y$ direction. Panels~(a) and (b)
depict situations where S is a nontopological superfluid (NSF).
Panel~(c) corresponds to the topological-transition point where
$h=h_\mathrm{c}\equiv\sqrt{\mu^2 + \big|\Delta^{(\mathrm{S})}
\big|^2}$, and panel~(d) is for S being a topological superfluid
(TSF). Blue curves show the energy dispersion of Andreev edge
states as a function of 2D-wave-vector component $k_y$ parallel
to the interface, arising from solution of the characteristic
equations (\ref{eq:NSFsecAA}) [panels (a) and (b)] and
(\ref{eq:TSFsecAA}) [panel (d)]. Black curves indicate the
minimum-energy bound of the quasiparticle-excitation continuum
calculated from the $2\times 2$-projected-theory dispersions
(\ref{eq:lowEdisp}). The arbitrary units $k_0$ and $E_0$ for
wave number and energy are related via $E_0=\hbar^2 k_0^2/(2
m)$. Results shown are for $\lambda k_0/E_0 = 0.2$, $\big|
\Delta^{(\mathrm{S})}\big|/E_0 = 0.1$, $\mu/E_0 = 1$ in all
panels and $h/E_0 = 0.3$ in (a), $0.9$ in (b), $\sqrt{1.01}$ in
(c) and $1.02$ in (d).}
\end{figure}

The characteristic equation (\ref{eq:NSFsecAA}) for edge states
of the NSF is formally analogous to the general expression
\begin{align}\label{eq:pWaveJunct}
& \cos\left[ \theta^{(\mathrm{S'})} + \sgn(E)\,\,
\gamma^{(\mathrm{S'})}\, \vartheta^{(\mathrm{S'})}_{k_y} \right]
\,\, \cos\left[\theta^{(\mathrm{S})} - \sgn(E)\,\,
\gamma^{(\mathrm{S})}\, \vartheta^{(\mathrm{S})}_{k_y} \right]
\nonumber \\[0.2cm] & \hspace{0.2cm} = \frac{T}{2-T} \left\{
\cos\left(\varphi^{(\mathrm{S'})} - \varphi^{(\mathrm{S})}
\right) + \sin\left[ \theta^{(\mathrm{S'})} + \sgn(E)\,
\gamma^{(\mathrm{S'})}\, \vartheta^{(\mathrm{S'})}_{k_y} \right]
\,\, \sin\left[ \theta^{(\mathrm{S})} - \sgn(E)\,
\gamma^{(\mathrm{S})}\, \vartheta^{(\mathrm{S})}_{k_y} \right]
\right\}
\end{align}
for the equation determining Andreev-bound-state energies at a
Josephson (SS$^\prime$) junction between two spinless chiral
\textit{p}-wave superfluids~\cite{Ho1984,Matsumoto1999,
Barash2001,Kwon2004,Samokhin2012}. In Eq.~(\ref{eq:pWaveJunct}),
the superscript $R\in \{\mathrm{S}, \mathrm{S'}\}$ labels
quantities from the left (S) and right (S$^\prime$) sides of the
junction, respectively, $\gamma^{(R)}=\pm 1$ are the chiralities
of the \textit{p}-wave pair potentials, $\varphi^{(R)}$ is the
pair-potential phase in region $R$, and $0\le T\le 1$ is the
junction's transparency for particle transmission. See
Appendix~\ref{app:chiPwave} for a detailed derivation. Setting
$\varphi^{(\mathrm{S'})} - \varphi^{(\mathrm{S})}\to 0$ and
$T/(2-T)\to \lambda^2 k^{(\mathrm{S})}_{\mathrm{F}\uparrow}
k^{(\mathrm{S})}_{\mathrm{F}\downarrow}/(2 h^2)$ in
(\ref{eq:pWaveJunct}), together with an adjusted definition of
$\vartheta_{k_y}$, yields (\ref{eq:NSFsecAA}). Physically, this
makes sense, as the opposite-spin subsystems of the NSF, each
constituting an effective realization of a spinless
chiral-\textit{p}-wave superfluid, are coupled in a way that is
similar to a Josephson junction via the finite spin-orbit
coupling $\lambda\ne 0$. As it is well-known~\cite{Barash2001,
Kwon2004} that the energy of the Andreev bound state at
chiral-\textit{p}-wave SS$^\prime$ junctions with
$\varphi^{(\mathrm{S'})} - \varphi^{(\mathrm{S})}=0$ is finite
for $k_y\ll k_\mathrm{F}$ [see also Fig.~\ref{fig:ABS} in
Appendix~\ref{app:chiPwave}], we expect the same for the NSF
edge states. Plotting solutions of (\ref{eq:NSFsecAA}) confirms
our expectation; see Figs.~\ref{fig:ISenergy}(a) and
\ref{fig:ISenergy}(b). As the minimum of the subgap-excitation
energy occurs close to (albeit not exactly at) $k_y=0$, it is
useful to note the analytical result
\begin{equation}\label{eq:NSFky0ABS}
|E(k_y=0)| \approx \frac{\lambda^2 k^{(\mathrm{S})}_{\mathrm{F}
\uparrow} k^{(\mathrm{S})}_{\mathrm{F}\downarrow}}{h^2}\, \big|
\Delta^{(\mathrm{S})}\big| \equiv \frac{\lambda
k^{(\mathrm{S})}_{\mathrm{F}\uparrow}}{h}\,
\Delta^{(\mathrm{S})}_\downarrow \quad .
\end{equation}
Thus while the Andreev-edge-state energy for the NSF is finite,
it is small because we are focusing on the situation where
$\lambda k^{(\mathrm{S})}_{\mathrm{F}\uparrow} \ll h$.
Furthermore, it vanishes together with
$\Delta^{(\mathrm{S})}_\downarrow$ at the topological
transition; see Fig.~\ref{fig:ISenergy}(c).

Turning to the solution of the characteristic equation
(\ref{eq:TSFsecAA}) for the Andreev-edge-state energy of the
I-TSF boundary, we find
\begin{equation}\label{eq:Majorana}
E(k_y) = - \left[ 1 + \frac{\lambda^2 \big(
k^{(\mathrm{S})}_{\mathrm{F}\uparrow}\big)^2}{2 h^2} \right]
\frac{\Delta^{(\mathrm{S})}_\uparrow}
{k^{(\mathrm{S})}_{\mathrm{F}\uparrow}}\, k_y \equiv - \left[ 1
+ \frac{\lambda^2 \big( k^{(\mathrm{S})}_{\mathrm{F}\uparrow}
\big)^2}{2 h^2} \right] \frac{\lambda\big|\Delta^{(\mathrm{S})}
\big|}{h}\, k_y \quad .
\end{equation}
Apart from the $\lambda^2$-dependent correction in brackets,
the result (\ref{eq:Majorana}) coincides with the dispersion of
the familiar~\cite{Honerkamp1998,Matsumoto1999,Furusaki2001,
Stone2004,Mizushima2008,Fu2008,Sauls2011} Majorana edge mode of
a chiral-\textit{p}-wave superconductor, with pertinent
parameters from the spin-$\uparrow$ sector of the TSF. (See the
discussion in the last paragraph of Appendix~\ref{app:chiPwave},
as well as the entire Appendix~\ref{app:Majorana}, for relevant
background information.) A specific example is shown in
Fig.~\ref{fig:ISenergy}(d). As terms of order $\mathcal{O}(k_y^2
/k_{\mathrm{F}\uparrow}^2)$ are neglected as part of the Andreev
approximation, we do not resolve nonlinear corrections to the
topological-edge-state dispersion that should become important
when it approaches the quasiparticle-excitation continuum. The
TSF-edge mode with dispersion (\ref{eq:Majorana}) corresponds
to the IS-interface Andreev bound state indicated by the solid
green line in Fig.~\ref{fig:ISSprime}(a): it is a quasiparticle
with dominant-spin-$\uparrow$ character and propagates in
negative-$y$ direction along the interface.

Our analytical results reproduce salient features seen in
numerically obtained excitation spectra for our system of
interest, e.g., those shown in insets of Fig.~7 from
Ref.~\cite{Holst2022}. While nontopological edge states in
superconducting spin-orbit-coupled nanowires have been studied
extensively (see Ref.~\cite{Kell2012} as one of the seminal
works and Refs.~\cite{Pan2020,Prada2020} for comprehensive
overviews), the existence of the nontopological edge state at
the I-NSF boundary in a 2D system has not been discussed so
far. Unlike the topologically protected (Majorana) edge state of
the I-TSF system, the subgap excitation at the NSF edge is not
robust against perturbations. Nevertheless, as it influences the
low-energy physics of the I-NSF system, spectroscopic
techniques~\cite{Vale2021,Flensberg2021} should be able to
detect it and distinguish it from the Majorana mode.

\section{Results \& discussion II: Bound states at the TSF-NSF
interface}\label{sec:resInter}

Technically, the TSF-NSF interface exhibits all the hallmarks of
a complicated multi-band Josephson junction. However,
considerable simplifications would arise under the assumption of
perfect decoupling for the majority-spin (spin-$\uparrow$) and
minority-spin (spin-$\downarrow$) sectors within the TSF and NSF
phases. In this limit, the interface becomes equivalent to a
configuration of two parallel Josephson junctions: one between
the chiral-\textit{p}-wave superfluids realized in the
spin-$\uparrow$ sectors of the TSF and NSF regions, and the
other one between those of the spin-$\downarrow$ sectors.
The chiral-\textit{p}-wave Josephson junction formed by the
spin-$\uparrow$ degrees of freedom from the TSF and NSF regions
is different from previously considered types~\cite{Ho1984,
Matsumoto1999,Barash2001,Kwon2004,Samokhin2012} because the
order-parameter magnitudes on opposite sides of the interface
are different. Furthermore, with the spin-$\downarrow$ degrees
of freedom being Zeeman-energy-quenched in the TSF, the TSF-NSF
interface can be expected to act as a wall for the
spin-$\downarrow$ chiral-\textit{p}-wave superfluid in the NSF.

While certainly being attractive for its simplicity, the
perfect-decoupling limit potentially fails to describe reality.
As described in detail in Sec.~\ref{sec:formal}, the
opposite-spin sectors of a spin-orbit-coupled polarized 2D Fermi
superfluid are not completely independent, and this residual
coupling can have important physical consequences. For example,
application of the perfect-decoupling limit to the I-NSF
interface would result in the expectation that it hosts a
helical edge mode, consisting of the two counter-propagating
chiral edge modes arising from individual chiral-\textit{p}-wave
superfluids realized with the spin-$\uparrow$ and
spin-$\downarrow$ sectors. That this is not the case is known
from numerical studies and general topological
considerations~\cite{Sato2010,Sato2009a}. Our detailed
calculations described in Sec.~\ref{sec:ISform} and results
discussed in Sec.~\ref{sec:resSurf} elucidate how the properties
of the I-NSF interface are manifestly shaped by the coupling
between opposite-spin sectors of the NSF. It is therefore
necessary to investigate whether and how the conclusions derived
for the TSF-NSF interface within the na{\"\i}ve picture that
assumes perfectly decoupled spin sectors are altered. This has
been the purpose of our careful analysis performed in
Sec.~\ref{sec:SSPform}, and we now discuss our obtained results.

\begin{figure}[t]
\centerline{%
\includegraphics[width=0.9\textwidth]{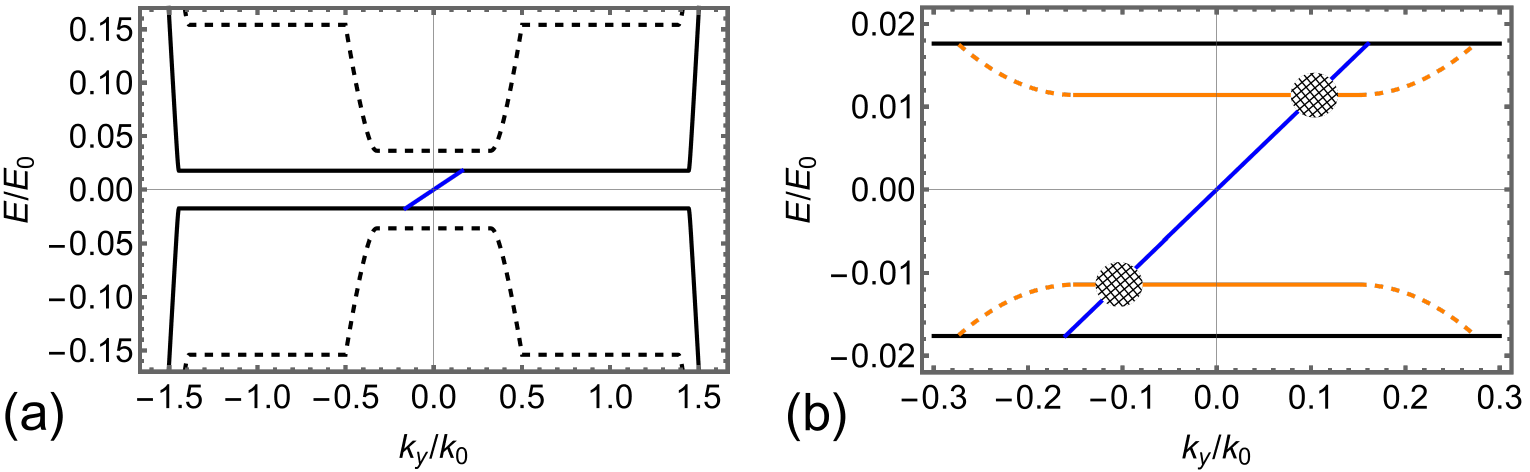}
}%
\caption{Andreev bound states and quasiparticle-continuum edges
at the  TSF (S) -- NSF (S$^\prime$) interface. The
interface-localised Andreev bound state with Majorana character
is realised in the minority-spin (spin-$\downarrow$) sector.
Panel (a) shows its energy dispersion from
Eq.~(\ref{eq:MajoranaSSp}) as the diagonal blue line. The solid
(dashed) black curves indicate the boundary of the
quasiparticle-excitation continua in the TSF (NSF) region
calculated from the $2\times 2$-projected-theory dispersions
(\ref{eq:lowEdisp}). Panel~(b) is a zoom-in to the gap region,
including additionally the Andreev bound states from the
majority-spin (spin-$\uparrow$) sector. Solid orange curves plot
the $k_y$-independent dispersions of Eq.~(\ref{eq:sUpABS}) with
$\varphi^{(\mathrm{S'})} - \varphi^{(\mathrm{S})} = 3\pi/4$.
Dashed continuations indicate the expected shape of these
dispersions at larger $k_y$, which lie beyond the approximations
of our theory. The hatched circles cover regions where the
Andreev-bound-state energies of both spin sectors are not
well-separated and results obtained within the
uncoupled-spin-sector approach no longer apply. The parameters
chosen correspond to the SS'-interface of 
Fig.~\ref{fig:ISSprime}(c) where the bulk quasiparticle
excitation gap is smaller in the TSF compared to the NSF. The
parameter values are $\lambda k_0/E_0 = 0.2$, $\mu/E_0 = 1.0$,
$h/E_0 = 1.05$, $\big|\Delta^{(\mathrm{S})}\big|/E_0 = 0.06$ and
$\big|\Delta^{(\mathrm{S'})}\big|/E_0 = 0.58$. The arbitrary
wave-number and energy scales $k_0$ and $E_0$ are related via
$E_0 = \hbar^2 k_0^2/(2m)$.}
\label{fig:SSpEnergy}
\end{figure}

Modifications to the NSF's spin-$\downarrow$ edge mode at the
TSF-NSF interface arising from its coupling to the
spin-$\uparrow$ sector are captured by the characteristic
equation (\ref{eq:SSpDownL2}). This expression closely
resembles the characteristic equation (\ref{eq:TSFsecAA}) for
the TSF edge. However, the parametric smallness of
$k^{(\mathrm{S'})}_{\mathrm{F}\downarrow}$ that is inherent in
our envisioned realization of the TSF-NSF hybrid system via a
spatially varying \textit{s}-wave pair potential [see
Eqs.~(\ref{eq:DeltaX}) and (\ref{eq:TSFtoNSF}), as well as the
associated discussion in Sec.~\ref{sec:intro}] implies that the
$\lambda^2$-dependent correction term in (\ref{eq:SSpDownL2})
should be neglected within the Andreev approximation, yielding
the characteristic equation (\ref{eq:TSFsecSSpAA}) obtained for
the completely decoupled spin-$\downarrow$ sector. Solving
(\ref{eq:TSFsecSSpAA}) to find the energy dispersion for the
edge state localised at the $\mathrm{SS'}$ junction yields
\begin{equation}\label{eq:MajoranaSSp}
E_\downarrow(k_y) = \frac{\Delta^{(\mathrm{S'})}_\downarrow}
{k^{(\mathrm{S'})}_{\mathrm{F}\downarrow}}\, k_y \equiv
\frac{\lambda\big| \Delta^{(\mathrm{S'})}\big|}{h}\, k_y \quad .
\end{equation}
Figure~\ref{fig:SSpEnergy}(a) shows a plot of $E(k_y)$, together
with the minimum-quasiparticle-excitation-energy bounds of the S
and S$^\prime$ regions, for an illustrative set of parameters.
With $\Delta_\mathrm{c} - \big|\Delta^{(\mathrm{S})}\big| =
\big|\Delta^{(\mathrm{S'})}\big| - \Delta_\mathrm{c} = 0.80\,
\Delta_\mathrm{c}$ being satisfied [see
Eq.~(\ref{eq:junctDeltaCri}) for the definition of the
junction's critical \textit{s}-wave pair-potential magnitude
$\Delta_\mathrm{c}$], this system's S and S$^\prime$ regions
represent well-developed TSF and NSF phases, respectively. A
zoom-in to the low-energy, small-$k_y$ region is shown in
Fig.~\ref{fig:SSpEnergy}(b), including also interface-localized
Andreev bound states from the spin-$\uparrow$ sector.

The general form of Eq.~(\ref{eq:MajoranaSSp}) is analogous to
that of the result (\ref{eq:Majorana}), without its small
$\lambda^2$-dependent correction, obtained when solving the
I-TSF-edge characteristic equation (\ref{eq:TSFsecAA}). Thus, to
leading order in small spin-orbit-coupling magnitude, the same
type of linear-in-$k_y$ dispersion indicative of Majorana
excitations in a chiral-\textit{p}-wave superfluid emerges for
both the I-TSF edge mode and the spin-$\downarrow$ TSF-NSF
interface mode. However, these modes exhibit crucial
differences. Firstly, the opposite overall signs of their
dispersions indicate that the propagation direction of the
SS$^\prime$-interface edge state is opposite to that of the IS
edge state, as illustrated by Fig.~\ref{fig:ISSprime}.
Furthermore, the TSF edge state is associated with
quasiparticles having dominant spin-$\uparrow$, i.e.,
majority-spin character, while the TSF-NSF interface edge state
is formed by minority-spin (spin-$\downarrow$) quasiparticles.
Finally, the familiar form of the Majorana-edge-mode velocity
for a chiral-\textit{p}-wave superfluid that is given by the
\textit{p}-wave pair-potential magnitude divided by the Fermi
wave vector applies in both cases. However, because of the way
that $\Delta_\sigma^{(R)}$ depends also on $k_{\mathrm{F}
\sigma}^{(R)}$ [see Eq.~(\ref{eq:pAbbrev})], the Majorana-mode
velocity at the SS$^\prime$ interface turns out to be a measure
of the \textit{s}-wave pair-potential magnitude in the
S$^\prime$ part, while the velocity of the IS-edge Majorana mode
depends on the \textit{s}-wave pair-potential magnitude of the S
region. Within our particular scenario where the relation $\big|
\Delta^{(\mathrm{S})}\big| < \big|\Delta^{(\mathrm{S}^\prime)}
\big|$ holds [inferred from Eq.~(\ref{eq:TSFtoNSF}); see also
Fig.~\ref{fig:ISSprime}], the SS$^\prime$ interface edge mode
will be faster than the IS edge mode. Thus, while the TSF region
depicted in Fig.~\ref{fig:ISSprime} has the required edge modes
at its boundaries with nontopological vacuum (I) and a
nontopological superfluid (NSF), the edge-mode properties are
not necessarily determined solely by TSF-region parameters.

The excitation spectrum of the chiral-\textit{p}-wave Josephson
junction formed by the spin-$\uparrow$ degrees of freedom from
the TSF and NSF regions turns out to be unaffected by coupling
to the spin-$\downarrow$ sector as long as the energy of the
Majorana edge mode realized in the spin-$\downarrow$ sector is
well-separated from that of the spin-$\uparrow$ bound state.
(See Sec.~\ref{sec:SSPupC} for a detailed discussion.) 
With this condition, majority-spin Andreev-bound-state energies
are solutions of the characteristic equation
(\ref{eq:sUpSSpABS}), yielding the explicit expression
\begin{equation}\label{eq:sUpABS}
E_{\uparrow\eta} (k_y) = \eta\,\, \frac{\lambda\,
k_{\mathrm{F}\uparrow}^{(\mathrm{SS'})}}{h}\,\sqrt{\big|
\Delta^{(\mathrm{S})}\big|\big|\Delta^{(\mathrm{S'})}\big|}\,\,
\frac{\cos\left[ \left(\varphi^{(\mathrm{S'})} -
\varphi^{(\mathrm{S})}\right)/2\right]}{\sqrt{1 +
\frac{G}{\sin^2\left[ \left(\varphi^{(\mathrm{S'})} -
\varphi^{(\mathrm{S})}\right)/2\right]}}} \quad .
\end{equation}
In Eq.~(\ref{eq:sUpABS}), $\eta \in \{+, -\}$ distinguishes the
positive-energy and negative-energy branches, $G$ is defined in
Eq.~(\ref{eq:GPdef}), and $\arccos\big( \big|
\Delta^{(\mathrm{S})}\big|/\big|\Delta^{(\mathrm{S'})}\big|\big)
< \varphi^{(\mathrm{S'})} - \varphi^{(\mathrm{S})} < 2\pi -
\arccos\big( \big|\Delta^{(\mathrm{S})}\big|/\big|
\Delta^{(\mathrm{S'})}\big|\big)$ is required for a bound state
to exist. Thus the difference in order-parameter magnitudes on
opposite sides of the interface manifests in $G\ne 0$ and the
restriction on the range of order-parameter phase differences
across the junction. The $k_y$-independence of (\ref{eq:sUpABS})
will be modified at larger $k_y$ by corrections
$\mathcal{O}(k_y^2)$ that we have neglected within the Andreev
approximation. Figure~\ref{fig:SSpEnergy}(b) plots $E_{\uparrow
\eta}(k_y)$ from Eq.~(\ref{eq:sUpABS}) for a particular
realization of the TSF-NSF interface together with the
dispersion for the spin-$\downarrow$-sector Majorana mode
[$E_\downarrow(k_y)$ from Eq.~(\ref{eq:MajoranaSSp})].

\section{Conclusions}
\label{sec:concl}

We have studied theoretically the subgap excitations emerging at
boundaries and interfaces of the polarized 2D Fermi superfluid
subject to spin-orbit coupling and \textit{s}-wave attraction.
While the specific form of 2D-Rashba-type~\cite{Bychkov1984,
Bihlmayer2015,Manchon2015} spin-orbit coupling has been assumed
in all of our calculations, our conclusions apply more generally
also to other $\kk$-linear types such as
2D-Dirac~\cite{Winkler2015} and
2D-Dresselhaus~\cite{Dresselhaus1955,Eppenga1988}. We juxtapose
the properties of edge excitations at the vacuum boundary with
those of Andreev bound states localized at the interface between
topological and nontopological phases of the polarized
spin-orbit-coupled 2D Fermi superfluid. See
Fig.~\ref{fig:ISSprime} for a schematic overview of our system
of interest.

The theoretical formalism employed for deriving the
Andreev-bound-state energy dispersions and Nambu-spinor forms
utilizes and extends the accurate effective description of 
mixed-spin quasiparticle excitations in terms of their dominant
spin-projected Nambu-spinor amplitudes~\cite{Brand2018}.
Section~\ref{sec:formal} provides an introduction to this
formalism that is tailored to this work's objectives and
contains relevant mathematical details of its extension. Our
approach provides a systematic platform for mapping the
low-energy physics of the spin-orbit-coupled polarized 2D Fermi
gas with \textit{s}-wave pairing to the excitation spectra of
effective chiral-\textit{p}-wave superfluids realized within
majority-spin (spin-$\uparrow$) and minority-spin
(spin-$\downarrow$) subspaces. (See Fig.~\ref{fig:WaveVecs} for
an illustration.) It is a crucial feature of our theory that the
intrinsic coupling between opposite-spin sectors is carefully
treated to ensure realistic physical predictions. The methods
developed here enable a versatile treatment of Bogoliubov
quasiparticles in superfluids with multiple coupled degrees of
freedom.

While the edge excitations of chiral-\textit{p}-wave
superfluids~\cite{Honerkamp1998,Matsumoto1999,Furusaki2001,
Stone2004,Mizushima2008,Fu2008,Sauls2011} and noncentrosymmetric
superconductors~\cite{Eschrig2012,Holst2022} have been studied
extensively, much less attention has been focused on the
interface between topological and nontopological superfluids.
Our present work sheds new light on the Andreev-bound-state
properties for both these venues: the insulator-superfluid (IS)
boundary and the superfluid-superfluid (SS$^\prime$) junction.
In particular, we demonstrate how the boundary of the 2D
\textit{s}-wave Fermi superfluid in the nontopological phase
establishes a Josephson junction between the
chiral-\textit{p}-wave superfluids realized in the system's
spin-$\uparrow$ and spin-$\downarrow$ sectors, with the
spin-orbit coupling determining the effective junction
transparency. This is epitomized by the analogous formal
structure of Eqs.~(\ref{eq:NSFsecAA}) and (\ref{eq:pWaveJunct}).
We also elucidate corrections to the Majorana-edge-mode
dispersion in the topological phase [Eq.~(\ref{eq:Majorana})]
arising from the coupling between opposite-spin sectors. Most
importantly, we establish the emergence of a Majorana mode at
the interface between topological and nontopological phases of
the spin-orbit-coupled polarized 2D Fermi superfluid and
elucidate its distinct physical properties. In particular, its
velocity depends on the \textit{s}-wave pair potential of the
nontopological part of the system and is thus different from
that of the Majorana edge mode of the topological phase with
vacuum. [Compare Eqs.~(\ref{eq:Majorana}) and
(\ref{eq:MajoranaSSp}).] Having different Majorana-mode
velocities at the IS boundary and the SS$^\prime$ interface
constitutes an experimental signature for the different origin
of the Majorana excitation in both locations. Another
characteristic that is different between these two Majorana
modes is their opposite spin, as illustrated in
Fig.~\ref{fig:ISSprime}.

Tunneling spectroscopy has been proposed as an experimental tool
for identifying Bogoliubov-quasiparticle excitations with
Majorana character realized in semiconductor-superconductor
hybrid structures~\cite{Beenakker2016} by their vanishing
excitation energy~\cite{Flensberg2021} or the equal magnitude of
their particle and hole admixtures~\cite{Cao2022}. Spectroscopic
techniques for measuring the energies and wave functions of
quasiparticle excitations in cold-atom systems are also
available~\cite{Vale2021}. Specifically, spatially resolved rf
spectroscopy~\cite{Shin2007} can be adapted to measure the local
density of states, thus serving as a cold-atom analog of
tunneling spectroscopy~\cite{Jiang2011a}. The utility of this
technique for probing Majorana edge modes in a 2D topological
Fermi superfluid has already been pointed out~\cite{Liu2012}, and
we expect it to be similarly useful for facilitating
experimental detection of the Majorana excitation at an
interface between coexisting topological and nontopological
superfluids. Spin-resolved measurements~\cite{Vale2021} would be
able to confirm our prediction that the interface-localized
Majorana mode has minority-spin character, in contrast to the
Majorana excitation from the majority-spin sector present at the
vacuum boundary.

Current interest in the spin-orbit-coupled polarized 2D Fermi
superfluid has been fuelled largely by the system's topological
phase simulating features of a chiral-\textit{p}-wave
superfluid. The microscopic basis for such behavior is a
good separation between majority-spin and minority-spin
excitations. We have used our theoretical formalism for a
systematic exploration of the effect that any residual coupling
between the opposite-spin subsystems has on Andreev bound states
at the IS vacuum boundary or an SS$^\prime$ interface. At the
SS$^\prime$ interface between topological and nontopological
phases, the emergent Majorana mode in the minority-spin sector
and the chiral-\textit{p}-wave junction established in the
majority-spin sector turn out to be robust as long as the
Andreev-bound-state energies in both subsystems are
well-separated. Future work may explore the interplay between
these two in the strong-coupling regime. It would also be useful
to study the potential for hybridization between the Majorana
modes localized at the IS boundary and SS$^\prime$ interface
when the S-region width $L$ becomes small (see
Fig.~\ref{fig:ISSprime}). The fact that these two modes are
from opposite-spin sectors leads us to expect that hybridization
is even more strongly suppressed than between Majorana modes
at opposite vacuum boundaries of a topological superfluid (i.e.,
an ISI system).

In closing, we briefly address certain features of real physical
systems that have not been accounted for in our formalism.
\textit{(i)~Finite-temperature ($T>0$) effects.\/} The fact that
the Andreev bound states considered in this work have subgap
energies provides them with some protection against thermal
fluctuations, as long as $k_\mathrm{B} T$ is smaller than their
energy separation from the nearest quasiparticle continuum. This
criterion limits the $k_y$ range of any dispersive mode, but it
is least restrictive for chiral Majorana excitations as the
latter disperse linearly around a zero-energy state. The thermal
energy will also influence the size of the region indicated by
hatched circles in Fig.~\ref{fig:SSpEnergy}(b) where a more
sophisticated theoretical description of the TSF-NSF interface
bound states is needed. More broadly, proper $T$-dependent values
will have to be used for all parameters entering quantitative
predictions presented in this work. In this context, we expect
the thermal suppression of pair-potential magnitudes to result
in the most significant adjustments. \textit{(ii) Effect of a
trapping potential.\/} Especially in cold-atom realizations of a
TSF-NSF hybrid system, smoothly varying external potentials
introduce spatial inhomogeneity. Our results apply most directly
to situations with optical-box potentials~\cite{Navon2021}, 2D
versions of which have been used for trapping fermionic
atoms~\cite{Hueck2018} and realising an ideal Josephson
junction~\cite{Luick2020}. In the more common case of
harmonically trapped 2D fermion systems (see, e.g.,
Ref.~\cite{Dyke2011}), the particle density, chemical potential
and \textit{s}-wave pair potential acquire smooth spatial
variations on the scale of the trap's harmonic-oscillator
length. As the Andreev bound states are localized at a system
boundary or the TSF-NSF interface within much shorter length
scales, their qualitative features can be expected to be largely
unaffected by the trap. Quantitative predictions made within
this work should also still apply if local values of, e.g., the
\textit{s}-wave pair potential are used. \textit{(iii) Dipolar
interactions} are expected to stabilize chiral-\textit{p}-wave
pairing~\cite{Cooper2009,Levinsen2011,Fedorov2016}. Current
interest in ultracold dipolar gases~\cite{Baranov2008,
Baranov2012,Chomaz2023} motivates extension of our study to
describe Andreev bound states at boundaries of such systems.

\section*{Acknowledgements}
We thank Philip Brydon for helpful discussions.


\paragraph{Funding information}
This work was partially supported by the
\href{https://search.crossref.org/funding?q=501100009193&from_ui=yes}{\sf Marsden Fund}
of New Zea\-land (contract nos.\ VUW1713 and MAU2007) from
government funding managed by the Royal Society Te Ap\=arangi.

\begin{appendix}

\section{Andreev bound states at
chiral-\textit{p}-wave-superfluid junctions}
\label{app:chiPwave}

If the residual weak coupling between opposite-spin degrees of
freedom in a spin-orbit-coupled polarized 2D Fermi superfluid
with \textit{s}-wave attraction is neglected, then the system is
effectively split into separate spin-$\uparrow$ and
spin-$\downarrow$ parts that constitute 2D
chiral-\textit{p}-wave superfluids with opposite chirality. A
spin-conserving implementation of the SS$^\prime$ interface then
establishes separate junctions for the spin-$\uparrow$ and
spin-$\downarrow$ sectors. In this Appendix, we discuss the
general properties of spinless chiral-\textit{p}-wave superfluid
junctions to inform the understanding of the TSF-NSF hybrid
system considered in the main part of the Article in the limit
where opposite-spin subsystems are completely decoupled. 

We consider an SS$^\prime$ hybrid system consisting of two
spinless 2D chiral-\textit{p}-wave superfluids occupying the
half-spaces $x>0$ and $x<0$, respectively. Within a given
subspace $R\in \{\mathrm{S}, \mathrm{S'}\}$, the superfluid is
described by a $2\times 2$ BdG equation
\begin{equation}\label{eq:pBdG}
\begin{pmatrix}
\frac{\hbar^2}{2 m^{(R)}}\, \kkop^2 + \nu^{(R)} - \mu &
\vek{\Delta}^{(R)} \cdot \frac{\kkop}{k_\mathrm{F}^{(R)}}
\\[0.2cm] \big[ \vek{\Delta}^{(R)}\big]^\ast \cdot
\frac{\kkop}{k_\mathrm{F}^{(R)}} & -\left[ \frac{\hbar^2}{2
m^{(R)}}\, \kkop^2 + \nu^{(R)} - \mu \right] \end{pmatrix}
\begin{pmatrix} u^{(R)}(\rr)\\[0.3cm] v^{(R)}(\rr) \end{pmatrix}
= E \begin{pmatrix} u^{(R)}(\rr) \\[0.3cm] v^{(R)}(\rr)
\end{pmatrix} \quad ,
\end{equation}
with $k_\mathrm{F}^{(R)} = \sqrt{2 m^{(R)} \big( \mu - \nu^{(R)}
\big)/\hbar^2}$. All system paramters, including the band-bottom
shift $\nu^{(R)}$ and the chiral-\textit{p}-wave order-parameter
vector $\vek{\Delta}^{(R)}\equiv \big( \Delta^{(R)}_x ,
\Delta^{(R)}_y \big) = \big( \Delta^{(R)} , \gamma^{(R)} i\,
\Delta^{(R)} \big)$, are constants within region $R$, and
$\gamma^{(R)} = +1$ ($-1$) encodes the positive (negative)
chirality. Translational invariance along the junction motivates
the separation \textit{Ansatz\/}
\begin{equation}\label{eq:1DsepAns}
\begin{pmatrix} u^{(R)}(\rr) \\[0.1cm] v^{(R)}(\rr)
\end{pmatrix} = \begin{pmatrix} u(x) \\[0.1cm] v(x)
\end{pmatrix} \, \ee^{i k_y y} \quad ,
\end{equation}
which transforms the 2D BdG equation (\ref{eq:pBdG}) into an
effective 1D BdG equation for motion perpendicular to the
interface,
\begin{equation}\label{eq:chiP1DBdG}
\begin{pmatrix}
\frac{\hbar^2}{2}\, \hat{k}_x\, \frac{1}{m(x)} \, \hat{k}_x +
\frac{\hbar^2 k_y^2}{2 m(x)} + \nu(x) - \mu &
\frac{1}{k_\mathrm{F}(x)} \big\{ \Delta_x(x)\, , \, \hat{k}_x
\big\} + \Delta_y(x) \, \frac{k_y}{k_\mathrm{F}(x)} \\[5pt]
\frac{1}{k_\mathrm{F}(x)} \big\{ \Delta_x^\ast(x)\, , \,
\hat{k}_x \big\} + \Delta_y^\ast(x) \, \frac{k_y}{k_\mathrm{F}
(x)} & -\left[ \frac{\hbar^2}{2}\, \hat{k}_x\, \frac{1}{m(x)}\,
\hat{k}_x + \frac{\hbar^2 k_y^2}{2 m(x)} + \nu(x) - \mu \right]
\end{pmatrix} \begin{pmatrix} u(x) \\[5pt] v(x) \end{pmatrix} =
E \begin{pmatrix} u(x) \\[5pt] v(x) \end{pmatrix} .
\end{equation}
Here $\{ A \, , B \} \equiv (A B + B A)/2$ denotes the
symmetrized product of two operators, and we have allowed for
all relevant parameters to be piecewise-constant;
\begin{subequations}
\begin{align}
m(x) &= m^{(\mathrm{S})}\, \Theta(-x) + m^{(\mathrm{S'})}\,
\Theta(x) \quad , \\
\nu(x) &= \nu^{(\mathrm{S})}\, \Theta(-x) + \nu^{(\mathrm{S'})}
\, \Theta(x) \quad , \\
k_\mathrm{F}(x) &= k_\mathrm{F}^{(\mathrm{S})}\, \Theta(-x) +
k_\mathrm{F}^{(\mathrm{S'})}\, \Theta(x) \quad , \\
\Delta_j(x) &= \Delta_j^{(\mathrm{S})}\, \Theta(-x) +
\Delta_j^{(\mathrm{S'})}\, \Theta(x) \quad .
\end{align}
\end{subequations}
Assuming $\mu - \nu^{(R)} > 0$, the general \textit{Ansatz\/}
for the $x$-dependent part of the Nambu spinor is
\begin{align}\label{eq:BdGansatz}
\begin{pmatrix} u(x) \\[5pt] v(x) \end{pmatrix} &= \left[
a_{+ -}^{(\mathrm{S})}\, \begin{pmatrix} u^{(\mathrm{S})}_{+ -}
\\[5pt] v^{(\mathrm{S})}_{+ -} \end{pmatrix}\, \ee^{i\,
k^{(\mathrm{S})}_{+ -}\,x} \,\, +\,\,  a_{- +}^{(\mathrm{S})}\,
\begin{pmatrix} u^{(\mathrm{S})}_{- +} \\[5pt]
v^{(\mathrm{S})}_{- +} \end{pmatrix}\, \ee^{i\,
k^{(\mathrm{S})}_{- +}\,x} \right] \Theta(-x) \nonumber \\ &
\hspace{3cm} + \left[ a_{+ +}^{(\mathrm{S'})}\, \begin{pmatrix}
u^{(\mathrm{S'})}_{+ +} \\[5pt] v^{(\mathrm{S'})}_{+ +}
\end{pmatrix}\, \ee^{i\, k^{(\mathrm{S'})}_{+ +}\,x} \,\, +\,\,
a_{- -}^{(\mathrm{S'})}\, \begin{pmatrix} u^{(\mathrm{S'})}_{-
-} \\[5pt] v^{(\mathrm{S'})}_{- -} \end{pmatrix}\, \ee^{i\,
k^{(\mathrm{S'})}_{- -}\,x} \right] \Theta(x) \quad ,
\end{align}
with the wave numbers and Nambu-spinor entries given explicitly
as
\begin{subequations}\label{eq:pIntPar}
\begin{align}\label{eq:kxChiP}
k^{(R)}_{\tau\alpha} &= \alpha\,\, \sqrt{\left(
k_{\mathrm{F}}^{(R)}\right)^2 - k_y^2 - 2\left( \frac{m^{(R)}
\big|\Delta^{(R)}\big|}{\hbar^2 k_\mathrm{F}^{(R)}} \right)^2
+\tau\, \frac{2 m^{(R)}}{\hbar^2} \, \sqrt{E^2 - \big|
\Delta^{(R)}\big|^2 \left[ 1 - \left(\frac{m^{(R)}\big|
\Delta^{(R)}\big|}{\hbar^2 \big( k_\mathrm{F}^{(R)} \big)^2}
\right)^2\right]}} \,\, , \\[5pt] \label{eq:uAmpChiP}
u_{\tau\alpha}^{(R)} &= \exp\left( i \varphi^{(R)}\right) \,\,
\frac{k_{\tau\alpha}^{(R)} +\gamma^{(R)} i\, k_y}{\sqrt{\big(
k_{\tau\alpha}^{(R)}\big)^2 + k_y^2}} \nonumber \\ &
\hspace{1.5cm} \times \sqrt{\frac{1}{2E}\left( E - \frac{m^{(R)}
\big|\Delta^{(R)}\big|^2}{\hbar^2 \big( k_\mathrm{F}^{(R)}
\big)^2} + \tau\, \sqrt{E^2 - \big|\Delta^{(R)}\big|^2 \left[ 1
- \left( \frac{m^{(R)}\big|\Delta^{(R)}\big|}{\hbar^2 \big(
k_\mathrm{F}^{(R)}\big)^2}\right)^2\right]}\right)}\,\, ,\\[5pt]
v_{\tau\alpha}^{(R)} &=\sgn(E)\,\,\,\,\sqrt{\frac{1}{2E}\left( E
+ \frac{m^{(R)} \big|\Delta^{(R)}\big|^2}{\hbar^2 \big(
k_\mathrm{F}^{(R)}\big)^2} - \tau\, \sqrt{E^2 - \big|
\Delta^{(R)}\big|^2 \left[ 1 - \left( \frac{m^{(R)}\big|
\Delta^{(R)}\big|}{\hbar^2 \big(k_\mathrm{F}^{(R)}\big)^2}
\right)^2\right]}\right)}\,\, .
\end{align}
\end{subequations}
The matching conditions at the SS$^\prime$ interface read
\begin{subequations}\label{eq:matchP}
\begin{align}
\begin{pmatrix} u(x) \\[5pt] v(x) \end{pmatrix}_{x\to 0^-} -
\begin{pmatrix} u(x) \\[5pt] v(x) \end{pmatrix}_{x\to 0^+} &= 0
\,\, , \\[0.2cm] \label{eq:SSpDeriv} \left[ \frac{1}{m(x)}\,
\frac{d}{dx} \begin{pmatrix} u(x)\\[5pt] v(x) \end{pmatrix}
\right]_{x\to 0^-} - \left[ \frac{1}{m(x)}\, \frac{d}{dx}
\begin{pmatrix} u(x) \\[5pt] v(x) \end{pmatrix}\right]_{x\to
0^+} &= \begin{pmatrix} 0 & -i \kappa \\[5pt] i \kappa^\ast & 0
\end{pmatrix} \begin{pmatrix} u(0) \\[5pt] v(0) \end{pmatrix}
\,\, ,
\end{align}
with the generally complex quantity
\begin{equation}
\kappa = \frac{\big|\Delta^{(\mathrm{S})}\big|}{\hbar^2
k_\mathrm{F}^{(\mathrm{S})}} \, \exp\left(
i\varphi^{(\mathrm{S})}\right) - \frac{\big|
\Delta^{(\mathrm{S'})}\big|}{\hbar^2
k_\mathrm{F}^{(\mathrm{S'})}}\, \exp\big(
i\varphi^{(\mathrm{S'})} \big) \quad .
\end{equation}
\end{subequations}
The $\kappa$-dependent terms on the right-hand side of
Eq.~(\ref{eq:SSpDeriv}) have been neglected in previous
works~\cite{Ho1984,Matsumoto1999,Barash2001,Kwon2004,
Samokhin2012} as their formalism employed the Andreev
approximation~\cite{Andreev1964} from the start. Applying the
conditions (\ref{eq:matchP}) to the \textit{Ansatz\/}
(\ref{eq:BdGansatz}) yields the characteristic equation
\begin{align}\label{eq:genSecChiP}
0 =& \left[ \left( \frac{k_{++}^\mathrm{(S')}}{m^\mathrm{(S')}}
- \frac{k_{+-}^\mathrm{(S)}}{m^\mathrm{(S)}} \right) \left(
\frac{k_{-+}^\mathrm{(S)}}{m^\mathrm{(S)}} -
\frac{k_{--}^\mathrm{(S')}}{m^\mathrm{(S')}}\right) - |\kappa|^2
\right] \left( u_{++}^\mathrm{(S')}\, u_{+-}^\mathrm{(S)}\,
v_{-+}^\mathrm{(S)}\, v_{--}^\mathrm{(S')} +
v_{++}^\mathrm{(S')}\, v_{+-}^\mathrm{(S)}\, u_{-+}^\mathrm{(S)}
u_{--}^\mathrm{(S')} \right) \nonumber \\[5pt]
&+ \left[ \left( \frac{k_{++}^\mathrm{(S')}}{m^\mathrm{(S')}} -
\frac{k_{-+}^\mathrm{(S)}}{m^\mathrm{(S)}} \right) \left(
\frac{k_{--}^\mathrm{(S')}}{m^\mathrm{(S')}} -
\frac{k_{+-}^\mathrm{(S)}}{m^\mathrm{(S)}} \right) + |\kappa|^2
\right] \left( u_{+-}^\mathrm{(S)}\, v_{++}^\mathrm{(S')}\,
v_{-+}^\mathrm{(S)}\, u_{--}^\mathrm{(S')} + v_{+-}^\mathrm{(S)}
\, u_{++}^\mathrm{(S')}\, u_{-+}^\mathrm{(S)}\,
v_{--}^\mathrm{(S')} \right) \nonumber \\[5pt]
&- \left( \frac{k_{++}^\mathrm{(S')}}{m^\mathrm{(S')}} -
\frac{k_{--}^\mathrm{(S')}}{m^\mathrm{(S')}} \right) \left(
\frac{k_{-+}^\mathrm{(S)}}{m^\mathrm{(S)}} -
\frac{k_{+-}^\mathrm{(S)}}{m^\mathrm{(S)}} \right) \left(
u_{++}^\mathrm{(S')}\, v_{+-}^\mathrm{(S)}\,
v_{-+}^\mathrm{(S)}\, u_{--}^\mathrm{(S')} +
v_{++}^\mathrm{(S')}\, u_{+-}^\mathrm{(S)}\, u_{-+}^\mathrm{(S)}
\, v_{--}^\mathrm{(S')} \right) \nonumber \\[5pt]
&- \left( \frac{k_{++}^\mathrm{(S')}}{m^\mathrm{(S')}} -
\frac{k_{--}^\mathrm{(S')}}{m^\mathrm{(S')}} \right) \left(
u_{+-}^\mathrm{(S)}\, v_{-+}^\mathrm{(S)} - v_{+-}^\mathrm{(S)}
\, u_{-+}^\mathrm{(S)} \right) \left( \kappa^\ast\,
u_{++}^\mathrm{(S')}\, u_{--}^\mathrm{(S')} + \kappa\,
v_{++}^\mathrm{(S')} v_{--}^\mathrm{(S')} \right) \nonumber
\\[5pt] &- \left( \frac{k_{-+}^\mathrm{(S)}}{m^\mathrm{(S)}} -
\frac{k_{+-}^\mathrm{(S)}}{m^\mathrm{(S)}} \right) \left(
u_{++}^\mathrm{(S')}\, v_{--}^\mathrm{(S')} -
v_{++}^\mathrm{(S')}\, u_{--}^\mathrm{(S')} \right) \left(
\kappa^\ast \, u_{+-}^\mathrm{(S)}\, u_{-+}^\mathrm{(S')} +
\kappa \, v_{+-}^\mathrm{(S)}\, v_{-+}^\mathrm{(S)} \right)
\end{align}
for the interface-localized Andreev-bound-state energies.

Equation (\ref{eq:genSecChiP}) applies generally for all
situations where $\mu-\nu^{(\mathrm{S,S'})}>0$. Leaving a more
complete analysis of the solutions to future work, we focus on
the case when the Andreev approximation~\cite{Andreev1964} is
applicable. Within this approach, terms $\propto\kappa$ in
(\ref{eq:genSecChiP}) can be neglected, and
\begin{subequations}\label{eq:AapproxP}
\begin{align}
k^{(R)}_{\tau\alpha} &\approx \alpha\, k_\mathrm{F}^{(R)} +
\alpha\tau\, \frac{m^{(R)}}{\hbar^2 k_\mathrm{F}^{(R)}}\,
\sqrt{E^2 - \big( \Delta^{(R)} \big)^2} \,\, , \\[5pt]
u_{\tau\alpha}^{(R)} &\approx \alpha\,\, \exp\left\{ i \left[
\varphi^{(R)} + \frac{\tau\, \sgn(E)}{2}\, \theta^{(R)} +
\alpha\, \gamma^{(R)}\, \vartheta^{(R)}_{k_y} \right] \right\}
\,\, \sqrt{\frac{\big|\Delta^{(R)}\big|}{2 |E|}} \,\, , \\[5pt] 
v_{\tau\alpha}^{(\mathrm{R})} &\approx \sgn(E)\,\, \exp\left[ -i
\frac{\tau\, \sgn(E)}{2}\, \theta^{(R)} \right] \,\,
\sqrt{\frac{\big|\Delta^{(R)}\big|}{2 |E|}}\,\, ,
\end{align}
\end{subequations}
with the phase angle $\theta^{(R)} = \arccos\big(|E|/
\big|\Delta^{(R)}\big|\big)$, and $\vartheta^{(R)}_{k_y} =
\arcsin\big(k_y/k^{(R)}_\mathrm{F}\big)$ is reminiscent of a
quasiparticle's angle of incidence on the interface. Using the
approximations from Eqs.~(\ref{eq:AapproxP}) in the
characteristic equation (\ref{eq:genSecChiP}), the latter
simplifies to
\begin{align}\label{eq:simpPwaveSec}
& \cos\left[ \theta^{(\mathrm{S'})}+\theta^{(\mathrm{S})} +
\sgn(E) \left( \gamma^{(\mathrm{S'})}\,
\vartheta_{k_y}^{(\mathrm{S'})} - \gamma^{(\mathrm{S})}\,
\vartheta_{k_y}^{(\mathrm{S})}\right)\right] \nonumber \\[5pt]
&+\,\, \frac{Z^2}{1+Z^2}\, \cos \left[ \theta^{(\mathrm{S'})} -
\theta^{(\mathrm{S})} + \sgn(E) \left( \gamma^{(\mathrm{S'})}\,
\vartheta_{k_y}^{(\mathrm{S'})} + \gamma^{(\mathrm{S})}\,
\vartheta_{k_y}^{(\mathrm{S})}\right)\right] = \frac{1}{1+Z^2}\,
\cos\left( \varphi^{(\mathrm{S'})} - \varphi^{(\mathrm{S})}
\right) \,\, .
\end{align}
Equation~(\ref{eq:simpPwaveSec}) generalizes previously
obtained~\cite{Ho1984,Barash2001,Kwon2004,Samokhin2012} forms of
the characteristic equation for Andreev bound states at
chiral-\textit{p}-wave junctions to the situation where
$\theta^{(\mathrm{S})} \ne \theta^{(\mathrm{S'})}$, and it is
also a special case of characteristic equations derived for
bound states at general un\-con\-ven\-tion\-al-superconductor
junctions~\cite{Tanaka1996,Kashiwaya2000}. That the
interface-transparency parameter~\cite{Blonder1983}
\begin{equation}\label{eq:Zdef}
Z = \frac{1}{2} \left|
\sqrt{\frac{m^{(\mathrm{S})}\,
k_\mathrm{F}^{(\mathrm{S'})}}{m^{(\mathrm{S'})}\, 
k_\mathrm{F}^{(\mathrm{S})}}} - \sqrt{\frac{m^{(\mathrm{S'})}\,
k_\mathrm{F}^{(\mathrm{S})}}{m^{(\mathrm{S})}\,
k_\mathrm{F}^{(\mathrm{S'})}}}\right|
\end{equation}
enters via the combinations $T=1/(1+Z^2)$ and $R\equiv 1-T =
Z^2/(1+Z^2)$ that correspond, respectively, to the quantum
probabilities for single-particle transmission and reflection
through the interface is a well-known feature of Josephson
junctions~\cite{Furusaki1991,Beenakker1991,Bagwell1992}.
Application of addition theorems for trigonometric functions
transforms the characteristic equation (\ref{eq:simpPwaveSec})
into the form given in Eq.~(\ref{eq:pWaveJunct}).

Analytical solution of (\ref{eq:simpPwaveSec}) for the
Andreev-bound-state energy $E_\alpha(k_y)$ is possible in
certain limits. For basic-illustration purposes, we provide here
the result for $k_y=0$;
\begin{equation}\label{eq:ChiP-ABS}
E_\eta(k_y = 0) = \eta\, E_0 \left\{ \frac{\cos^2(\varphi/
2)}{1+Z^2} + \frac{1}{1+Z^2}\,\, \frac{F(\varphi)}{2 Z^2} \left[
\sqrt{1 - 4 Z^2\, \cos^2(\varphi/2) \, \frac{G}{\left[
F(\varphi) \right]^2}} - 1 \right]\right\}^{\frac{1}{2}} .
\end{equation}
Here $\eta=\pm$ distinguishes two branches of bound-state
energies, and the parameters entering the expression
(\ref{eq:ChiP-ABS}) are
\begin{subequations}
\begin{align}\label{eq:E0Pdef}
E_0 &= \sqrt{\big|\Delta^{(\mathrm{S})}\big|\,
\big|\Delta^{(\mathrm{S'})}\big|} \quad , \\
\varphi &= \varphi^{(\mathrm{S'})} - \varphi^{(\mathrm{S})}
\quad , \\
F(\varphi) &= G + Z^2 + \sin^2(\varphi/2) \quad , \\
\label{eq:GPdef} G &= \frac{1}{4} \left( \sqrt{\frac{\big|
\Delta^{(\mathrm{S'})}\big|}{\big|\Delta^{(\mathrm{S})}\big|}} -
\sqrt{\frac{\big|\Delta^{(\mathrm{S})}\big|}{\big|
\Delta^{(\mathrm{S'})}\big|}}\right)^2 \quad , 
\end{align}
\end{subequations}
with $Z$ given in Eq.~(\ref{eq:Zdef}). When $(1 + 2 Z^2)\,\big|
\Delta^{(\mathrm{S})}\big| < \big|\Delta^{(\mathrm{S'})}\big|$,
bound states exist only for values of the phase difference
$\varphi$ across the junction that are within the restricted
range $\varphi_0 < \varphi < 2\pi - \varphi_0$,
with
\begin{equation}\label{eq:phi0P}
\varphi_0 = \arccos\left[ (1 + 2 Z^2)\, \frac{\big|
\Delta^{(\mathrm{S})}\big|}{\big|\Delta^{(\mathrm{S'})}\big|}
\right] \equiv \arccos \left[ \frac{1 + 2 Z^2}{\left( \sqrt{1+
G} + \sqrt{G} \right)^2} \right] \,\, .
\end{equation}
Without loss of generality, the definition (\ref{eq:phi0P}) of
$\varphi_0$ assumes $\big|\Delta^{(\mathrm{S})}\big|\le \big|
\Delta^{(\mathrm{S'})}\big|$. The fact that interface-localized
Andreev bound states can exist only within a restricted range of
$\varphi$ when pair-potential magnitudes are different on
opposites sides of the junction is also true for the
\textit{s}-wave case, which was overlooked in previous
work~\cite{Presilla2017}. Specializing (\ref{eq:ChiP-ABS}) to
the case $\big|\Delta^{(\mathrm{S})}\big|=\big|
\Delta^{(\mathrm{S'})}\big|\equiv |\Delta|$, corresponding to
having chiral-\textit{p}-wave pair-potential components of equal
magnitude on both sides of the junction, yields
\begin{equation}
E_\eta (k_y=0) \to \frac{\eta\, |\Delta| \, |\cos(\varphi/2)
|}{\sqrt{1 + Z^2}} \quad \mbox{for $G\to 0$} \,\,\, ,
\end{equation}
in agreement with previous results~\cite{Ho1984,Barash2001,
Kwon2004,Samokhin2012}.

\begin{figure}[t]
\centerline{%
\includegraphics[width=0.8\textwidth]{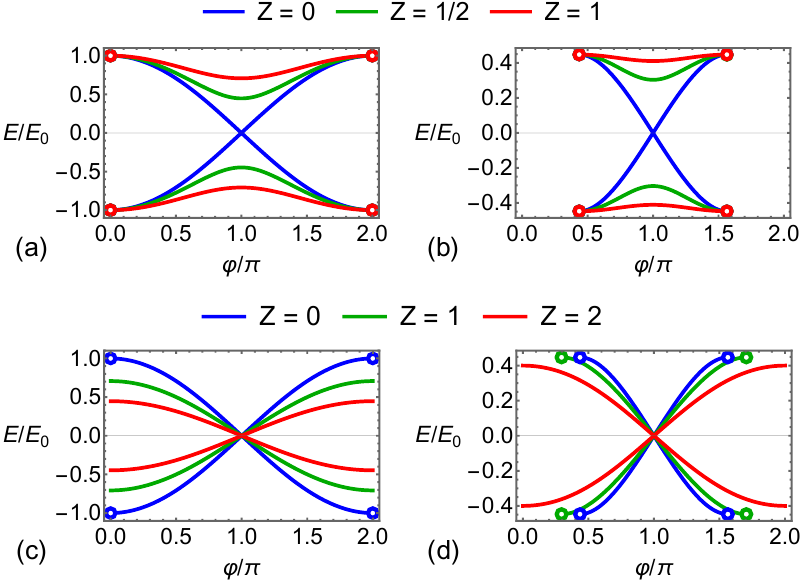}
}%
\caption{\label{fig:ABS}%
Energy $E$ of Andreev bound states at SS$^\prime$ interfaces,
obtained within the Andreev approximation. Results shown in
panels (a) and (b) [(c) and (d)] are for a junction between two
2D \textit{s}-wave superfluids with pair potentials $\big|
\Delta^{(\mathrm{S/S'})}\big|\exp\{(-/+)i\varphi/2\}$ [two 2D
chiral-\textit{p}-wave superfluids with pair potentials $\big|
\Delta^{(\mathrm{S/S'})}\big|\exp\{(-/+) i\varphi/2 \}\,\big\{
k_x +\gamma^{(\mathrm{S/S'})}\, i\, k_y\big\}/
k_\mathrm{F}^{(\mathrm{S/S'})}$ for the case with vanishing
wave-vector component parallel to the interface, i.e., $k_y=0$].
Situations with equal quasiparticle-gap magnitudes on both sides
of the SS$^\prime$ interface are depicted in panels (a) and (c),
corresponding to $G=0$. [See Eq.~(\ref{eq:GPdef}) for the
definition of $G$.] For comparison, panel (b) [(d)] is for
$G=4/5$ . The different curves within panels (a) and (b) [(c)
and (d)] are for values $Z=0$, $1/2$, and $1$ [$Z=0$, $1$ and
$2$] of the interface-transparency parameter $Z$ defined in
Eq.~(\ref{eq:Zdef}). Circles indicate end points of $\varphi$
ranges for which bound states exist. The energy unit $E_0$ is
defined in Eq.~(\ref{eq:E0Pdef}).}
\end{figure}

Panels (c) and (d) in Fig.~\ref{fig:ABS} show plots illustrating
parameter dependences of the $k_y=0$ Andreev-bound-state
energies at junctions between chiral-\textit{p}-wave
superfluids. Comparison with the results obtained for bound
states localized at the interface between \textit{s}-wave
superfluids [panels (a) and (b) of the same figure] reveals
distinctive properties of chiral-\textit{p}-wave junctions. One
particularly striking feature is the different effect normal
reflection (signified by $Z>0$) has on the energy
spectrum~\cite{Barash2001,Kwon2004}. For a junction of
\textit{s}-wave superfluids, the crossing at $\varphi=\pi$
becomes an anticrossing, while the energies at $\varphi =
\varphi_0$ and $2\pi-\varphi_0$ remain unaffected. In contrast,
for a chiral-\textit{p}-wave junction, the crossing at $\varphi=
\pi$ remains unaffected, while the magnitude of $\varphi_0$ gets
reduced and, eventually, the energies at $\varphi=0$ and $2\pi$
are suppressed by normal reflection. The physical origin for
this qualitatively different behavior lies in the additional
wave-vector-dependent phase shifts incurred for normal and
Andreev reflection off the interface with a
chiral-\textit{p}-wave superfluid, which shift the configuration
with constructive interference between normal and Andreev
reflection for interface-localized bound states from $\varphi=0$
for an \textit{s}-wave SS$^\prime$ junction to $\varphi=\pi$ for
the chiral-\textit{p}-wave case.

Analytical solutions of the characteristic equation
(\ref{eq:simpPwaveSec}) can also be obtained for a fully
transparent interface; i.e., $Z\to 0$. Assuming the velocity
parallel to the interface to have a magnitude not exceeding
$v_{y,\mathrm{max}}$ defined via $v_{y,\mathrm{max}}^2 =
v_\mathrm{F}^2\, \sqrt{G}/(\sqrt{1 + G} + \sqrt{G})$, where
$v_\mathrm{F} = \hbar\,k_\mathrm{F}^{(\mathrm{S})}/
m^{(\mathrm{S})}$ ($\equiv \hbar\,k_\mathrm{F}^{(\mathrm{S'})}/
m^{(\mathrm{S'})}$ for $Z=0$), we find
\begin{subequations}
\begin{equation}\label{eq:kyABS}
E_\eta(k_y) \to \frac{\eta\, E_0\, \cos\left(
\varphi_{k_y}^{(\eta)}/2\right)}{\sqrt{1 + \frac{G}{\sin^2\left(
\varphi_{k_y}^{(\eta)}/2\right)}}} \quad \mbox{for $Z\to 0$}
\quad ,
\end{equation}
with
\begin{equation}
\varphi_{k_y}^{(\eta)} = \varphi + \eta\left(
\gamma^{(\mathrm{S})} \, \vartheta_{k_y}^{(\mathrm{S})} -
\gamma^{(\mathrm{S'})} \, \vartheta_{k_y}^{(\mathrm{S'})}\right)
\end{equation}
\end{subequations}
and $\varphi_0 < \varphi_{k_y}^{(\eta)} < 2\pi - \varphi_0$.
Thus, $G\ne 0$ causes a linear $k_y$ dependence of the Andreev
bound-state energies for the same-chirality junction, which
normally (i.e., for $G=0$) is only a feature for
opposite-chirality junctions~\cite{Barash2001,Kwon2004}.

The situation where the SS$^\prime$ interface becomes a dividing
wall and the two chiral-\textit{p}-wave superfluids are
disconnected can be described by taking the limit $Z\to\infty$
in the characteristic equation (\ref{eq:simpPwaveSec}) or,
equivalently, by setting $T\to 0$ in (\ref{eq:pWaveJunct}). The
resulting characteristic equation can be expressed in the
factorized form
\begin{equation}\label{eq:SSpDivide}
\cos\left[ \theta^{(\mathrm{S'})} + \mathrm{sgn}(E)\,
\gamma^{(\mathrm{S'})}\, \vartheta^{(\mathrm{S'})}_{k_y} \right]
\,\, \cos\left[ \theta^{(\mathrm{S})} - \mathrm{sgn}(E)\,
\gamma^{(\mathrm{S})}\, \vartheta^{(\mathrm{S})}_{k_y} \right]
= 0 \quad ,
\end{equation}
which is satisfied if either one of the two individual cosine
factors appearing on the left-hand side of
Eq.~(\ref{eq:SSpDivide}) is separately equal to zero. Solution
of the two resulting conditions yield the Majorana edge
modes~\cite{Honerkamp1998,Matsumoto1999,Furusaki2001,Stone2004,
Mizushima2008,Fu2008,Sauls2011} associated with the left
boundary of the S$^\prime$ region and the right boundary of the
S region, having the respective dispersions
\begin{equation}\label{eq:DivideDisps}
E^{(\mathrm{S'})}(k_y) = \gamma^{(\mathrm{S'})}\,
\frac{|\Delta^{(\mathrm{S'})}|}{k_\mathrm{F}^{(\mathrm{S'})}}\,
k_y \quad , \quad E^{(\mathrm{S})}(k_y) = -\gamma^{(\mathrm{S})}
\, \frac{|\Delta^{(\mathrm{S})}|}{k_\mathrm{F}^{(\mathrm{S})}}\,
k_y \quad .
\end{equation}

\section{Majorana edge modes of chiral-\textit{p}-wave
superfluids}\label{app:Majorana}

The two independent solutions of the characteristic
equation~(\ref{eq:SSpDivide}), arising in the limit where the
$\mathrm{SS'}$ interface is a dividing wall, describe the edge
states propagating along the wall in the two separated S and
S$^\prime$ regions. Here we investigate the evanescent states
associated with the energy dispersions $E^{(\mathrm{S'})}(k_y)$
and $E^{(\mathrm{S})}(k_y)$ from Eq.~(\ref{eq:DivideDisps}).

To be specific, we start by considering the edge state of the
S$^\prime$ region. The $x$-dependent part of its Bogoliubov
spinor is given by the $x>0$ part of the \textit{Ansatz}
(\ref{eq:BdGansatz}), involving the complex wave-vector
components $k_{\tau\alpha}^{(\mathrm{S'})}$ and two-spinors
$\big(u_{\tau\alpha}^{(\mathrm{S'})},
v_{\tau\alpha}^{(\mathrm{S'})}\big)^T$ for which $\alpha\,\tau
\in\{+\, +, -\, -\}$. Using the Andreev-approximation
expressions given in Eqs.~(\ref{eq:AapproxP}) and neglecting
terms of magnitude $\mathcal{O}\big(\big[k_y/
k_\mathrm{F}^{(\mathrm{S'})}\big]^2\big)$ for consistency, we
find
\begin{align}
\begin{pmatrix} u_{++}^{(\mathrm{S'})} \\[5pt]
v_{++}^{(\mathrm{S'})} \end{pmatrix}_{E = E^{(\mathrm{S'})}
(k_y)}\,\, = \,\, \begin{pmatrix} u_{--}^{(\mathrm{S'})}
\\[5pt] v_{--}^{(\mathrm{S'})} \end{pmatrix}_{E =
E^{(\mathrm{S'})}(k_y)}\,\, \propto\,\, \begin{pmatrix}
\ee^{\frac{i}{2}\, \varphi^{(\mathrm{S'})}}\,\, (1+i) \\
\ee^{-\frac{i}{2}\, \varphi^{(\mathrm{S'})}}\,\, (1-i)
\end{pmatrix} \quad .
\end{align}
With this result, the properly normalized Bogoliubov-spinor wave
function (\ref{eq:1DsepAns}) for the edge state having
wave-vector component $k_y$ parallel to the interface in the
S$^\prime$ region is obtained as
\begin{align}\label{eq:MajStateSp}
\begin{pmatrix} u_{k_y}^{(\mathrm{S'})}(\rr) \\[5pt]
v_{k_y}^{(\mathrm{S'})}(\rr) \end{pmatrix} =
\frac{\Theta(x)}{\sqrt{2\pi l_\mathrm{coh}^{(\mathrm{S'})}}}\,\,
\sin\left( k_\mathrm{F}^{(\mathrm{S'})} x \right) \,\, \exp
\left( -\frac{x}{l_\mathrm{coh}^{(\mathrm{S'})}}\right)\,\,
\exp\left(i k_y y \right)\,\, \begin{pmatrix} \ee^{\frac{i}{2}
\, \varphi^{(\mathrm{S'})}}\,\, (1+i) \\ \ee^{-\frac{i}{2}\,
\varphi^{(\mathrm{S'})}}\,\, (1-i) \end{pmatrix} \quad ,
\end{align}
with the S$^\prime$ region's superfluid coherence length scale
$l_\mathrm{coh}^{(\mathrm{S'})}=\hbar^2
k_\mathrm{F}^{(\mathrm{S'})}/\big(m^{(\mathrm{S'})}\big|
\Delta^{(\mathrm{S'})}\big|\big)$. Performing a similar
calculation for the edge state propagating along the SS$^\prime$
interface in the S region yields the corresponding Bogoliubov
spinor
\begin{align}\label{eq:MajStateS}
\begin{pmatrix} u_{k_y}^{(\mathrm{S})}(\rr) \\[5pt]
v_{k_y}^{(\mathrm{S})}(\rr) \end{pmatrix} =
\frac{\Theta(-x)}{\sqrt{2\pi l_\mathrm{coh}^{(\mathrm{S})}}}\,\,
\sin\left( k_\mathrm{F}^{(\mathrm{S})} x \right) \,\, \exp
\left( \frac{x}{l_\mathrm{coh}^{(\mathrm{S})}}\right)\,\,
\exp\left(i k_y y \right)\,\, \begin{pmatrix} \ee^{\frac{i}{2}
\, \varphi^{(\mathrm{S})}}\,\, (1-i) \\ \ee^{-\frac{i}{2}\,
\varphi^{(\mathrm{S})}}\,\, (1+i) \end{pmatrix} \quad .
\end{align}

Both spinors (\ref{eq:MajStateSp}) and (\ref{eq:MajStateS})
satisfy
\begin{equation}
\left[ u_{k_y}^{(R)}(\rr) \right]^\ast = v_{-k_y}^{(R)}(\rr)
\quad ,
\end{equation}
which is the defining relation for a Majorana mode. This
is so because, in our system of interest, the particle-hole
conjugation operation $C$ is defined via~\cite{Elliott2015}
\begin{equation}
C\, \begin{pmatrix} u \\ v \end{pmatrix} = \begin{pmatrix} v^*
\\ u^* \end{pmatrix} \quad .
\end{equation}
Thus, it is possible to form a general superposition of
plane-wave edge states
\begin{align}
\Psi^{(R)}(\rr, t) &= \int_0^{k_\mathrm{F}^{(R)}} \frac{d k_y}{2
\pi}\,\, \left\{ f(k_y) \,\, \exp\left[ -\frac{i}{\hbar}\,
E^{(R)}(k_y)\, t \right] \,\, \begin{pmatrix} u_{k_y}^{(R)}(\rr)
\\[5pt] v_{k_y}^{(R)}(\rr)\end{pmatrix} \right. \nonumber
\\[5pt] & \hspace{5cm} \left. +\,\,\, \left[ f(k_y) \right]^*
\,\, \exp\left[ \frac{i}{\hbar}\, E^{(R)}(k_y)\, t \right] \,\,
\begin{pmatrix} u_{-k_y}^{(R)}(\rr) \\[5pt] v_{-k_y}^{(R)}(\rr)
\end{pmatrix} \right\}
\end{align}
that satisfies $C\, \Psi^{(R)}(\rr, t) = \Psi^{(R)}(\rr, t)$
and, therefore, describes a particle that is its own
particle-hole conjugate (i.e., \emph{antiparticle}).

\end{appendix}


\nolinenumbers

\end{document}